\documentclass{aa}
\usepackage{graphicx}
\usepackage{aalongtable}
\usepackage{longtable,lscape}
\usepackage{natbib}
\bibpunct{(}{)}{;}{a}{}{,}

\def\TMB {$T_{\rm mb}$}

\def\H0 {$H_{\rm o}$}

\def\solmass {\hbox{M$_{\odot}$}}

\def\colum {\hbox{$L_{\rm CO}$}}

\def\lsol {\hbox{L$_{\odot}$ }}
\def\lsols {\hbox{L$_{\odot}$}}

\def\solum {\hbox{L$_{\odot}$ }}
\def\solums {\hbox{L$_{\odot}$}}
\def\irlum {\hbox{$L_{\rm FIR}$ }}
\def\irlums {\hbox{$L_{\rm FIR}$}}

\def\numd {\hbox{$n\,({\rm H}_2$) }}

\def\kms {\hbox{${\rm km\,s}^{-1}$ }}
\def\kmss {\hbox{${\rm km\,s}^{-1}$}}

\def\percc {$\hbox{{\rm cm}}^{-3}$}    %cm-3
\def\arcsec {\hbox{$^{\prime\prime}$}}

\def\NH3 {\hbox{${\rm NH}_{3}$}}                  %NH3
\def\CH3C2H {\hbox{${\rm CH}_3{\rm C}_2{\rm H}$}} %CH3C2H
\def\FORM {\hbox{${\rm H}_2{\rm CO}$}}            %H2CO
\def\CH3OD {\hbox{${\rm CH}_3{\rm OD}$}}          %CH3OD
\def\HC15N {\hbox{${\rm HC}^{15}{\rm N}$}}        %HC15N
\def\HN13C {\hbox{${\rm HN^{13}C}$}}              %HN13C
\def\HCOP {\hbox{${\rm HCO}^+$}}                  %HCO+
\def\SO2 {\hbox{${\rm SO_{2}}$}}                  %SO2
\def\H2S {\hbox{${\rm H_{2}S}$}}                  %H2S
\def \la{\mathrel{\mathchoice   {\vcenter{\offinterlineskip\halign{\hfil
$\displaystyle##$\hfil\cr<\cr\sim\cr}}}
{\vcenter{\offinterlineskip\halign{\hfil$\textstyle##$\hfil\cr
<\cr\sim\cr}}}
{\vcenter{\offinterlineskip\halign{\hfil$\scriptstyle##$\hfil\cr
<\cr\sim\cr}}}
{\vcenter{\offinterlineskip\halign{\hfil$\scriptscriptstyle##$\hfil\cr
<\cr\sim\cr}}}}}
\def \ga{\mathrel{\mathchoice   {\vcenter{\offinterlineskip\halign{\hfil
$\displaystyle##$\hfil\cr>\cr\sim\cr}}}
{\vcenter{\offinterlineskip\halign{\hfil$\textstyle##$\hfil\cr
>\cr\sim\cr}}}
{\vcenter{\offinterlineskip\halign{\hfil$\scriptstyle##$\hfil\cr
>\cr\sim\cr}}}
{\vcenter{\offinterlineskip\halign{\hfil$\scriptscriptstyle##$\hfil\cr
>\cr\sim\cr}}}}}
\begin{document}
%
%\thesaurus{11.09.4, 11.14.1, 13.18.8}
%
\title{Dense gas in luminous infrared galaxies}
%\subtitle{I. The data}
%
\author{W.A.~Baan\inst{1}, C.~Henkel\inst{2}, A.F.~Loenen\inst{3,1},
A.~Baudry\inst{4} \and T.~Wiklind\inst{5}}
\offprints{W.A. Baan, \email{baan@astron.nl}}

\institute{
  ASTRON, Oude Hoogeveensedijk 4, 7991 PD Dwingeloo, The Netherlands
\and
    Max-Planck-Institut f{\"u}r Radioastronomie,
    Auf dem H{\"u}gel 69, D-53121 Bonn, Germany
\and
 Kapteyn Astronomical Institute, P.O. Box 800, 9700 AV Groningen, The Netherlands
\and
  Observatoire de l'Universit{\'e} de Bordeaux, F-33270 Floirac, France
\and
  Onsala Space Observatory, S-43900 Onsala, Sweden \& Space
  Telescope Science Institute, Baltimore, Maryland 21218, USA
}
\date{Received date 31/01/2007; accepted date 19/10/2007}
\authorrunning{Baan et al.}
\titlerunning{Dense gas in luminous infrared galaxies}
\abstract {} {Molecules that trace the high-density regions of the
interstellar medium have been observed in (ultra-)luminous
(far-)infrared galaxies, in order to initiate multiple-molecule
multiple-transition studies to evaluate the physical and chemical
environment of the nuclear medium and its response to the ongoing
nuclear activity.}
 {The HCN(1$-$0), HNC(1$-$0), \HCOP(1$-$0), CN(1$-$0) and CN(2$-$1),
CO(2$-$1), and CS(3$-$2) transitions were observed in sources
covering three decades of infrared luminosity including sources
with known OH megamaser activity. The data for the molecules that
trace the high-density regions have been augmented with data
available in the literature.}
 {The integrated emissions of high-density tracer molecules
show a strong relation to the far-infrared luminosity. Ratios of
integrated line luminosities have been used for a first order
diagnosis of the integrated molecular environment of the evolving
nuclear starbursts. Diagnostic diagrams display significant
differentiation among the sources that relate to initial
conditions and the radiative excitation environment. Initial
differentiation has been introduced between the FUV radiation
field in photon-dominated-regions and the X-ray field in
X-ray-dominated-regions. The galaxies displaying OH megamaser
activity have line ratios typical of photon-dominated regions. }
 {}

\maketitle
\section{Introduction}

Large amounts of high-density molecular gas play a crucial role in
the physics of (ultra-)luminous (far-)infrared galaxies
((U)LIRGs), giving rise to spectacular starbursts and possibly
providing the fuel for the emergence of active galactic nuclei
(AGN). A strong linear relation has been found between the far
infrared (\irlums) and HCN luminosity, the latter being an
indicator of the high density (\numd$\ga$10$^4$\,\percc) component
of the interstellar medium. This arises predominantly from the
nuclear region and indicates a close relation to the nuclear
starburst environment and the production of the FIR luminosity
\citep{GaoS2004a}. The molecular gas contributes a significant
fraction to the dynamical mass  of the central regions of FIR
galaxies \citep{1991A&ARv...3...47H, YoungScoville1991}.

Emissions of molecules that trace the high-density regions in
LIRGs and ULIRGs may serve to study the characteristics of the
extreme environments in the nuclear regions of starburst and AGN
nuclei. While much of the energy generation of FIR-luminous
galaxies has traditionally been attributed to AGN activity using
their optical signatures, the NIR and radio signatures suggest
significant starburst contributions \citep{GenzelEA1998,
BaanKlockner2006}. Circum-nuclear star formation in the densest
regions serves as a power source for the majority of ULIRGs such
that they even mimic the presence of an AGN
\citep{BaanKlockner2006}. The ULIRG population is also
characterized by OH megamasers (MM) \citep{Baan1989,
HenkelWilson1990} resulting from amplification of radio continuum
by FIR-pumped foreground gas. Few H$_2$CO MM are also found among
this population \citep{BaanHU1993, ArayaBH2004}.

The observed molecular emissions from the high-density components
could carry the signature of the nuclear region of the ULIRG and
its nuclear power generation. Recent studies consider the
HCN\,(1$-$0) versus CO\,(1$-$0) relation and its implications for
the star formation activity \citep{GaoS2004a, GaoS2004b}.
Discussions are ongoing in the literature about whether such
molecular emissions are accurate tracers of the high-density
medium at the site of star formation \citep{WuEvansEA2005,
GraciaCarpioGPC2006, Papadop2007}. It should be evident that only
multi-transition and multi-molecule (multi-dimensional) studies
can describe the multi-parameter environment of the nuclear
activity and further establish the connection between the
molecular signature and the nature of the nuclear energy source.
The translation or extrapolation from the characteristics of
individual star formation regions in our Galaxy to integrated
regions in nearby galaxies at larger distances or even to
unresolved nuclear regions in distant galaxies requires detailed
multi-species observations and comparisons combined with elaborate
theoretical modelling. The interpretation of molecular line data
has focused on the nature of the local excitation mechanisms in
the form of photon dominated regions (PDR) with far-ultraviolet
radiation fields and X-ray dominated regions (XDR)
\citep{1996A&A...306L..21L, MeijerinkS2005}. The molecular
properties of these two different regimes can be used as
diagnostic tools in interpreting the integrated properties of
galaxies.

This paper reflects a multi-dimensional approach to understanding
molecular line emissions that started in the early 90's, but is
only now coming to fruition. This study originally aimed to
establish a better link between molecular megamaser activity and
the molecular properties of luminous FIR galaxies but it has
changed into a general study of ULIRGs. Here we present data on
the molecular characteristics of normal and FIR galaxies across
the available range of \irlums. Data of the CO, HCN, HNC, \HCOP,
CN, and CS line emissions obtained for a representative group of
37 FIR-luminous and OH megamaser galaxies has been joined with the
additional data of 80 sources presented in the literature. Early
studies of the properties of CS \citep{MauersbergerHWH1989},
\HCOP\ \citep{NguyenEA1992}, and HNC \citep{HuttemeisterEA1995}
using small numbers of sources studied the relations between line
and FIR luminosities and the nature of the central engine. Our
molecular emission data show clear tendencies that cover a wide
regime of nuclear activity. This database forms the basis for a
first evaluation of the emission line properties and for further
study and modelling of the line properties.

\section{Characteristics of high-density tracers in galaxies}
\label{sec:HCNexcitation}

Extragalactic line ratios do not display the extreme ratios in the
Galaxy, because they are integrations of ensembles of different
types of clouds within one beam. Nevertheless, clear variations
reflect the response of the molecular emission to the varying
physical and chemical environments characterizing the regions of
nuclear activity.

{\bf HCN and HNC Molecules} - Steady-state excitation models show
that the abundance of HCN and its isotopomer HNC decreases with
increasing density and temperature \citep{SchilkeEA1992}. Values
for the HNC/HCN abundance ratio below unity are consistent with
steady-state models for higher-density gas at higher temperatures.
An HNC/HCN abundance ratio greater than unity suggests a rather
quiescent, low temperature gas. In dark clouds this ratio can be
more than 3.0 \citep{1984ApJ...287..681C} but in warm environments
in giant molecular clouds, the ratio may decrease to values of 0.5
to 0.2 \citep{1987ip...symp..561I}. In the Orion Hot Core the
observed HNC/HCN column density ratio can be as low as 0.013 with
values of 0.2 on the ridge \citep{SchilkeEA1992}.  The observed
HNC/HCN intensity ratio is found to vary significantly from 0.16
to 2 among galaxies with similar HCN/CO intensity ratios
\citep{AaltoPHC2002}, which may be connected with IR pumping
\citep{AaltoEA2007}. This suggests that HNC is not reliable as a
tracer of cold (10 K) high-density gas in LIR galaxies, as it is
in the Galaxy. HCN has thus been used to determine the gas masses
and the related star formation efficiencies \citep{GaoS2004b}.

Our exploratory LVG simulations suggest that, for a given density,
column density, and temperature, HCN(1$-$0) and HNC(1$-$0) have
similar intensities, but that HCO$^+$(1$-$0) is weaker. At higher
temperatures HNC tends to be selectively destroyed in favor of HCN
as long as the medium is not highly ionized. The canonical HNC/HCN
abundance ratio is 0.9, if the standard neutral production paths
are used \citep{GoldsmithEA1981}. In a highly ionized medium, low
HNC/HCN ratios would be prevented since HCNH$^+$ can form HCN and
HNC with equal probability \citep{AaltoPHC2002, WangEA2004}. The
passage of shocks may selectively destroy HNC and could
significantly reduce the steady state HNC/HCN abundance ratio
\citep{SchilkeEA1992}.

{\bf The HCO$^+$ Molecule} - \cite{HuttemeisterEA1995} suggest a
simplified picture where the molecules trace different gas
components: HCN in the warm and dense part, HNC in the cool and
slightly less dense part, and \HCOP\, in the slightly less dense
part, that is penetrated by cosmic rays and/or UV photons. The
HCN/\HCOP\ intensity ratio may vary from 0.5 to $\geq$4, which
indicates that the molecules probe distinct physical regions in
nearby galaxies \citep{NguyenEA1992}. Alternatively, there have to
be large-scale abundance differences and/or different types of
chemistry. The excitation of HCN and \HCOP\, in these galaxies
suggests low optical depth of less than unity and very low
excitation temperatures of $\leq$ 10\,K \citep{NguyenEA1992}. This
conclusion is confirmed by studies of many species in NGC\,4945
and other nearby galaxies \citep[e.g.,][]{WangEA2004}. A
multi-transition study of the high-density medium in Arp\,220 and
NGC\,6240 suggests that HCN would serve better as a tracer of
dense gas than \HCOP, which can be more sub-thermally excited and
as a radical avoids high densities \citep{GrevePGR2006}.

Shock propagation can strongly change the gas-phase chemistry such
that \HCOP\ is enhanced relatively to HCN in gas associated with
young supernova remnants \citep{DickinsonEA1980, Wootten1981}. The
shocked region of IC\,443 shows an enhancement of two orders of
magnitude in \HCOP relative to CO \citep{DickinsonEA1980}. This
enhancement of \HCOP\ would result from increased ionization due
to cosmic rays in the shocked dense material \citep{Wootten1981,
Elitzur1983}.

{\bf The CN Molecule} - The abundance of the cyanid radical CN is
enhanced in the outer regions of molecular clouds in PDR dominated
environments \citep{GreavesC1996, RodriguezEA1998}. At larger
depths the CN abundance declines rapidly \citep{JansenEA1995}. To
a significant extent CN is a photo-dissociation product of HCN and
HNC such that e.g. the CN/HCN abundance increases in the outer
regions of molecular clouds that are exposed to ultraviolet (FUV)
radiation \citep[see][]{RodriguezEA1998}. Therefore, the
CN/HCN\,(1$-$0) line intensity ratios may be used to probe the
physical and chemical conditions for Galactic and extragalactic
PDR sources.

In the Orion bar the CN abundance is enhanced at the inner edge of
the PDR, but it declines at larger depths \citep{JansenSHD1995}.
The Orion-KL region shows HCN\,(1$-$0) emission that is
significantly stronger than the HNC and CN\,(1$-$0) emission,
which is predicted for warm ($T$ $\ga$ 50\,K) gas
\citep{1997ApJ...482..245U}.

Using CN\,(1$-$0) and CN\,(2$-$1), as well as HCN\,(1$-$0)
measurements across the inner star-forming molecular disk in
M\,82, \cite{FuenteEA2005} find that $N_{\rm CN}$/$N_{\rm HCN}$
$\approx$ 5. They argue that such a high ratio is indicative of a
giant and dense PDR bathed in the intense radiation field of the
starburst and conclude that $A_V$ $\leq$ 5 in the clouds of M82,
because for optically thicker clouds the CN/HCN column density
ratio would be smaller than observed. The effect of enhanced
CN/HCN abundance and line temperature ratios is confirmed by
theoretical models of CN and HCN gas-phase chemistry
\citep{BogerS2005}. This suggests that the CN/HCN abundances ratio
in molecular clouds may  be used as probes of FUV and cosmic-ray
driven gas-phase chemistry, for a wide range of conditions.
\cite{BogerS2005} find that in dense gas, CN molecules are
characteristically and preferentially produced near the inner
edges of the \ion{C}{ii} zones in the PDRs. In addition, chemical
models \citep{1983ApJ...267..610K, 1996A&A...306L..21L} indicate
an enhancement of CN under high X-ray ionization conditions. The
prominence of HNC and CN relative to HCN has been considered a
measure of the evolutionary stage of the respective starburst
\citep{AaltoPHC2002}. This way the objects with strong HCN and
weak HNC and CN are dominated by a warm dense gas environment
early in the evolution.

{\bf The CS Molecule} - Multi-transition studies of CS in selected
Galactic clouds by \cite{SnellEA1984} show that there is
considerable density variation across their core regions, because
the lower (column) density outer regions contribute little to the
total CS emission. The cloud core is approximately two orders of
magnitude denser than the extended cloud, and the density rises
very steeply towards the core. LVG-modelling of the data suggests
a mean density of 5 to 10 $\times$ 10$^5$ cm$^{-3}$ with cloud
core masses ranging from 3 $\times$ 10$^5$ to 2 $\times$ 10$^3$
\solmass. The CS abundance varies considerably, peaking in the
clumped high-density medium. Chemical studies suggest that the
fractional abundance of CS is sensitive to both the abundances of
sulfur and oxygen \citep{GreadelLF1982}.

Although the CS lines may be brightest toward star-forming cores
in regions of high volume-density, ubiquitous low-level CS
emission, probably due to sub-thermally excited low-density gas,
is found to dominate the emission from the inner Galaxy
\citep{McQuinnEA2002} and its central region \citep{BallySWH1988}.
On the other hand, the study of the excitation state of CS in the
integrated nuclear environment of Arp\,220 and NGC\,6240 suggests
that CS probes the densest gas in the high-density medium better
than HCN or HCO$^+$ \citep{GrevePGR2006}.

\section{Observations}
\label{sec:observations}

\subsection{IRAM Pico Veleta (PV) observations}

Observations with the 30-m IRAM\footnote{Institute de
Radioastronomie Millim\'etrique is supported by INSU/CNRS
(France), the MPG (Germany), and the IGN (Spain).} telescope at
Pico Veleta (PV) (Spain) were carried out in Sept. 1994, May 1996,
Jan. 1997, and May 1997. At the frequencies of the $\lambda$
$\sim$ 3.4\,mm $J$=1$-$0 transitions of HCN, HNC, and \HCOP\ the
full width to half power beam size was 25$''$; for the $\lambda$ =
2.6 and 1.3\,mm CO $J$=1$-$0 and 2$-$1 transitions it was 21 and
13$''$. Typically, observations were started with the CO $J$=1$-$0
and 2$-$1 lines, then the lower frequency frontend was tuned to
HCN, HNC or \HCOP\, while keeping CO(2$-$1) as a sensitive
indicator of pointing quality.

During the first session in 1994, the 3\,mm SIS receiver had a 512
MHz bandwidth and a receiver temperature of 80$-$120 K below 100
GHz (image sideband rejection 5-10\,dB). Starting in 1996, the
renewed SIS receivers were single-sideband tuned with image
sideband rejections of 25$-$40\,dB. Receiver temperatures were
$T_{\rm rec}$(SSB) = 55$-$85\,K at a wavelength of 3.4\,mm,
100$-$140\,K at 2.6\,mm, 130\,K at 2.0\,mm, and 100$-$180\,K at
1.3\,mm. The observations were done under varied atmospheric
conditions, which were best for the runs in 1997. The system
temperatures during these sessions were 200$-$300\,K for the
80$-$90\,GHz range, 260$-$400\,K around 150\,GHz, and
700$-$1200\,K around 230\,GHz.

During the first session in 1994 we used the filterbank backend
(FB) in a $512 \times 1$\,MHz mode and the acousto-optical
spectrometer (AOS) in a $512 \times 1$\,MHz mode. During the 1996
and 1997 observations we used the AOS in $512 \times 2$\,MHz mode
in parallel with the FB in $256 \times 1$\,MHz mode. The spectral
resolution of the observations with 1\,MHz channels was 3.2\,\kms
at 115\,GHz and 1.6\,\kms at 230\,GHz.

The observing procedure included the use of the wobbling secondary
mirror in order to minimize baseline problems, which allowed
switching between source and reference positions placed
symmetrically at offsets of 4$'$ in azimuth. Each wobbler phase
had a duration of 2\,s. From continuum cross scans through nearby
pointing sources, we estimate the absolute pointing uncertainty to
be of $\la$3\arcsec. The positional alignment between the
receivers is estimated to be better than 2\arcsec. For all sources
single nuclear pointings were used.

The results of our observations are presented in units of main
beam brightness temperature \citep[\TMB;
e.g.][]{1989LNP...333..351D}. Using beam efficiencies of 0.78,
0.75, 0.69, and 0.52, we find main beam sensitivities of $S$/\TMB
= 4.86, 4.72, 4.62, and 4.44\,Jy/K at $\lambda$ $\sim$ 3.4, 2.6,
2.0, and 1.3\,mm. The calibration is affected by the pointing,
baseline stability, and receiver stability from session to
session. The overall accuracy of the data is estimated to be 15\%.
Stability of calibration between the four observing periods has
been ensured by using the same calibration sources during all of
the sessions. All data has been calibrated using the standard
chopper wheel method.

\subsection{SEST observations}

Two observing sessions were done at the Swedish/ESO\footnote{The
European Southern Observatory} Submillimeter Telescope (SEST) at
La Silla (Chile) during January 1995 and January 1997. System
performance during the first session prevented detection with
sufficient signal-to-noise ratio for the high-density lines of all
our target sources except for IR07160$-$6215. The data presented
in this paper have been predominantly obtained during the
successful second session.

The observations were done using two Low-Resolution AOS systems
with 1 GHz bandwidth having a total of 1440 channels for a single
polarization. The channel spacing was 1.4\,MHz, corresponding to
5.2 and 3.6 \kms\ at 80 and 115\,GHz. In 1997 the receiver
temperatures were $T_{\rm rec}$(SSB) = 100$-$180\,K and
120$-$140\,K in the 80$-$116\,GHz and 130$-$150\,GHz frequency
ranges, respectively. Observations were conducted under very
favorable atmospheric conditions and the actual system
temperatures were between 150 and 200\,K in the 80$-$90\,GHz band
and between 250 and 350\,K around 150\,GHz.

The target transitions for the observations were the HCN(1$-$0)
and \HCOP(1$-$0) transitions, which occur in a single bandpass,
and the HNC(1$-$0) transition. Parallel observations were done of
CS(3$-$2) using the other polarization. CO(1$-$0) observations
were done for sources without known earlier detections. Single
nuclear pointings were used for all sources.  Pointing
observations indicate that the absolute pointing uncertainty is of
the order of $\la$3\arcsec.

Calibration was done with a chopper wheel and using the $T_{\rm
A}^*$ intensity scale. The spectral data has been converted into a
main-beam brightness temperature \TMB, using a main-beam
efficiency of 0.75 at 86\,GHz, 0.70 at 115\,GHz and 0.60 at
150\,GHz. Conversion factors to flux units are 18.75, 18.90, and
20.50 Jy/K. The beam size is respectively 57$''$, 45$''$, and
34$''$.

\section{Results}
\label{sec:results}

The observational results have been presented in Online Figures
A.1 and A.2 and in Tables B.1 and B.2. In following evaluations,
the luminosity distance $D_{\rm L}$ has been taken from the
literature for sources with radial velocities below about 1500
\kms and determined from known emission line velocities or from
NED using $H_0$ = 72 \kms and $q_0$ = 0.5. The FIR luminosity
$L_{\rm FIR}$ has been determined from the 60 and 100$\mu m$ IRAS
fluxes using an IR-color correction factor 1.8
\citep{SandersM1996, SolomonEA1997}.

\indent{\bf IRAM Pico Veleta Results} - Our Pico Veleta 30-m
observations of 18 luminous FIR sources have resulted in 66
detections of emission lines for the following transitions: 18 for
CO(2$-$1), 11 for HCN(1$-$0), 8 for HNC(1$-$0), 11 for
\HCOP\,(1$-$0), 8 for CS(3$-$2), 6 for CN(1$-$0) and 4 for
CN(2$-$1).  The target sample for HCN, HNC, and \HCOP\ was
selected on the basis of FIR luminosity, the presence of OH MM
activity, or the strength of the CO $J$=1$-$0 line obtained
earlier with the 12-m NRAO telescope (Baan \& Freund 1992,
unpublished). Sources for observations of the cyano radical CN
were selected on the basis of the strength of the CO and HCN
lines. The CS(3$-$2) data has been obtained in parallel to the
observations at lower frequencies. High-density tracer transitions
in sources with published detections at the time of the
observations were not re-observed. Individual spectra toward the
nuclear positions obtained during our observations are presented
in \TMB\ units in Figs.\,A.1 and A.2. Results from gaussian
fitting are presented in Table B.1 after further conversion into
units of Jy \kmss.

{\bf SEST results} - The observations of 20 sources with the SEST
15-m telescope have resulted in a total of 52 new detections of
emission lines in 16 sources for the following transitions: 12 for
CO(1$-$0), 11 for HCN(1$-$0), 8 for HNC(1$-$0), 13 for
\HCOP\,(1$-$0), 7 for CS(3$-$2), and 1 for CN(1$-$0). Target
sources during these sessions have been selected on the basis of
the strength of known CO emission lines \citep{MirabelEA1990,
GarayMM1993}, and of FIR luminosity taken from the IRAS catalogue.
The source NGC\,660 is common between the Pico Veleta and SEST
data sets. High density tracer lines in sources with published
detections at the time of the observations were not re-observed.
Individual spectra toward the nuclear positions are presented in
Fig.\,A.3 in \TMB\ units and the results from gaussian fits are
given in Table B.1 after conversion into units of Jy \kmss. Target
sources without any line detection have been omitted from the
tables.

{\bf Detections from the Literature} - The detections of our
combined sample has been augmented with detections of molecular
lines tracing the high-density component reported in the
literature. In addition, we have used these detections to verify
and cross-calibrate our own detections where possible. Thus we
obtained detections of at least one of the HCN(1$-$0), HNC(1$-$0),
\HCOP(1$-$0), CN $N$=1$-$0 and $N$=2$-$1, and CS(3$-$2)
transitions for a total of 80 additional sources. A total of 167
individual published records have been used, which includes a
significant number of HCN detections from \cite{GaoS2004a}.

{\bf The Integrated Source Sample} - Our own sample of 37 sources
has been augmented with the data of 80 sources from the
literature. Table B.1 presents the available data on molecules
that trace low-density and high-density components of the medium.
The database contains a total number of 202 reliable detections of
high-density tracers, and 110 for CO\,(1$-$0) with 32 accompanying
CO\,(2$-$1) line detections. This (still) represents an incomplete
record for many of the sources. The database has  84 detections
plus 18 upper limits for HCN, 28 + 13 for HNC, 42 + 10 for \HCOP,
a total of 19 + 6 for CN(1$-$0), 13 + 7 for CN(2$-$1), and 17 + 20
for CS(3$-$2).  For 23 sources a complete set of HCN\,(1$-$0),
HNC\,(1$-$0) and \HCOP(1$-$0) tracer transitions were obtained and
10 sources have one out of three as an upper limit.

All data records for our integrated sample have been converted
from \TMB\ into units of Jy \kms for the PV and SEST results or
from antenna temperature for other telescopes. We also note that
some data in the literature has been obtained with various
single-dish instruments: the 30-m Pico Veleta telescope, the 15-m
SEST telescope, the 12-m NRAO telescope at Kitt Peak, the Onsala
20-m telescope, and the 14-m FCRAO telescope. The inclusion and
comparison of this data has been complicated by the incomplete
reporting of observing and calibration procedures. In many cases
the CO\,(1$-$0) and HCN(1$-$0) line strengths from different data
records have been used. General agreement was found for line data
from multiple source records with errors within 30\%.  Some
published line data records, particularly in \cite{GaoS2004a},
were not compatible with our data and other available data and
could not be used in the compilation.  As a result some line
ratios are also different from those in other publications because
of inconsistencies between our own data and some published data.
The source NGC\,660 was included in both our observing samples and
a comparison of the HCN and \HCOP\ lines shows agreement within
10\%. We also note that some of our own lower quality spectra and
their unreliable fits have not been used in the evaluation and
have been designated as upper limits in our data base in Tables 1
and 2.

{\bf Evaluating the Data} -  The collected data has been obtained
at different observing beams and different beam filling factors.
For a given telescope, the beam sizes at the observing frequencies
of HCN, HNC, and \HCOP\ are very similar and they differ by some
25\% from those of CN\,(1$-$0) and CO\,(1$-$0). Since the emission
of the high-density tracers is well-confined to the nuclear
regions of the galaxies, it will generally not fill the beam. On
the other hand, the CO\,(1$-$0) emission and possibly also the
CO\,(2$-$1) emission will be more extended and could fill the beam
for nearby galaxies. The extent of a typical CO\,(1$-$0) emission
region in ULIRGs varies between 2 and 3 kpc
\citep{DownesSolomon1998}; this would result in an underestimate
of the CO emission at radial velocities less than 2000 \kms for
the PV telescope and less than 1000 \kms for the SEST telescope.

Because of their similar frequencies, the line ratios of the
high-density tracers are not affected by different beam filling
factors. Their ratios with CO\,(1--0) would only be affected by an
underestimate of the total CO emission. The line intensity ratio
between CO\,(1--0) and the high density tracers used throughout
this paper will therefore be representative for more distant
galaxies, while they could be lower limits for nearby galaxies. We
will return to the effect of relative filling factors in the
evaluation of the CO line ratios below (Sec. \ref{sec:COratios}).

\begin{figure}[htbp]
\begin{center}
\includegraphics[width=7cm,clip]{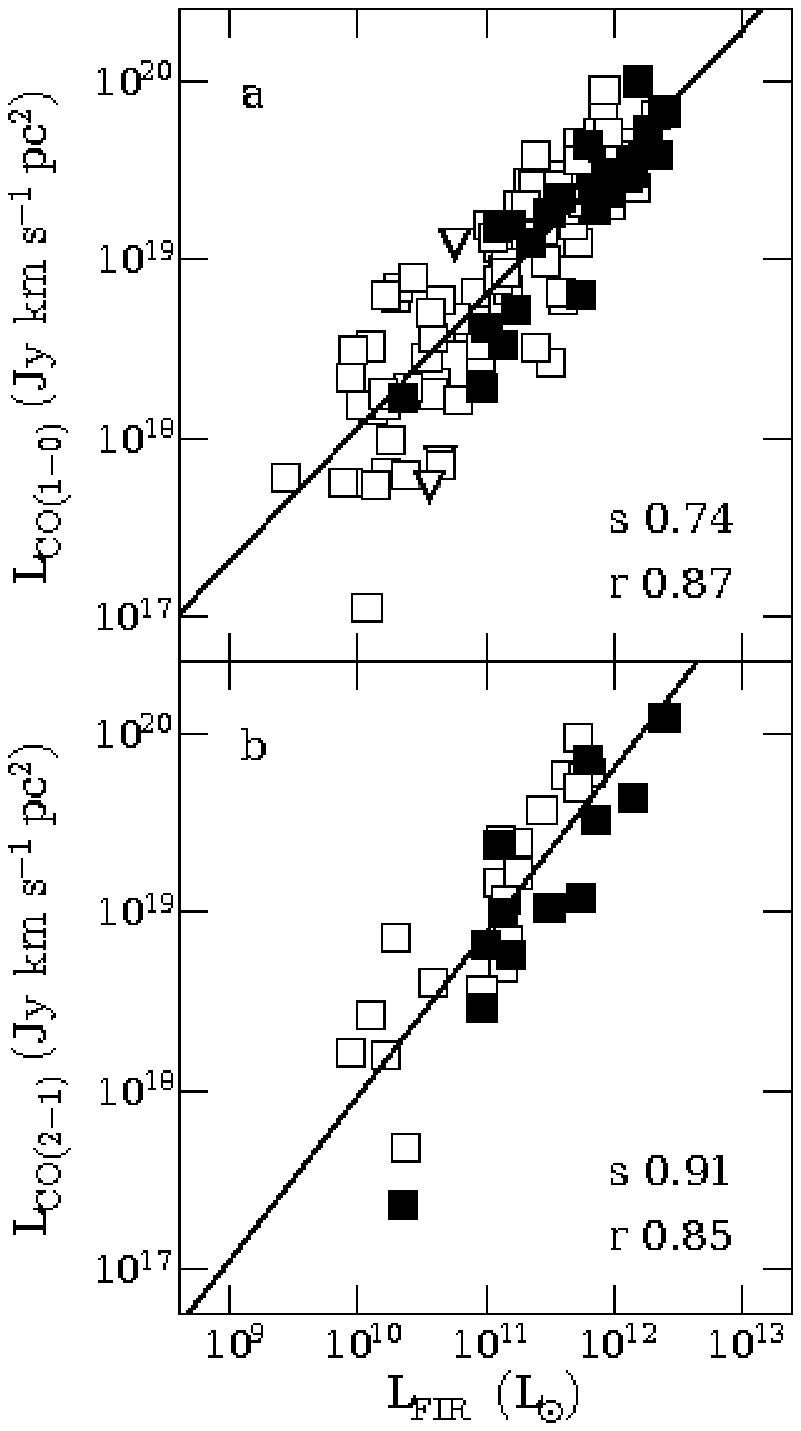}
\includegraphics[width=7cm,clip]{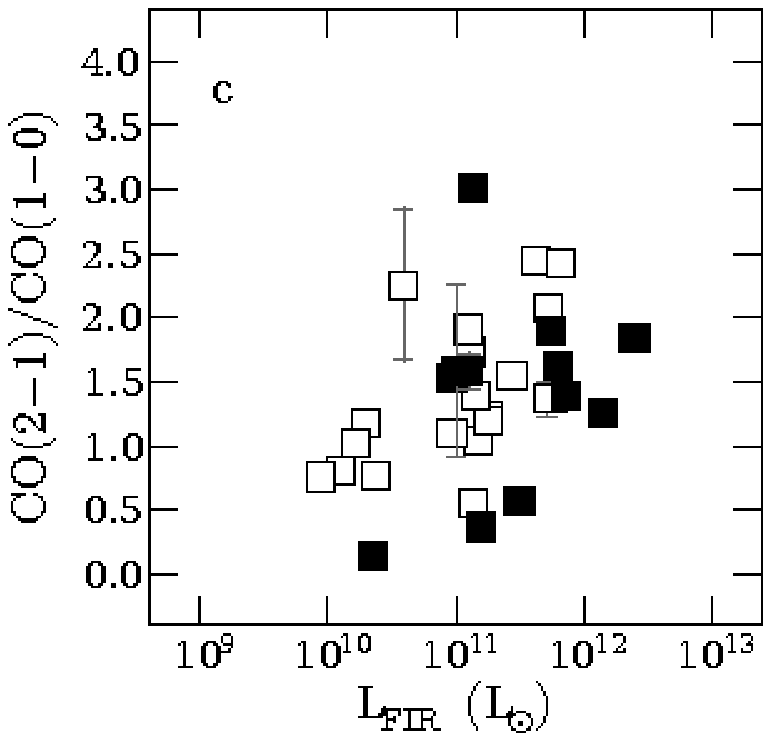}
\caption{Integrated line luminosities versus FIR luminosity for
the integrated sample of sources. a) CO\,(1$-$0) luminosity versus
FIR luminosity, b) CO\,(2$-$1) luminosity versus FIR luminosity,
and c) CO\,(2$-$1)/CO\,(1$-$0) luminosity ratio versus FIR
luminosity. There are 110 reliable data points for CO(1$-$0) and
32 for CO(2$-$1). The fitted slope ($s$) and the correlation
coefficient ($r$) are given in the lower right corner of each
frame and the fit does not take into account any displayed upper
limits. The data points in the diagrams are either squares for
reliable values or triangles for upper and lower limits. Filled
symbols indicate the 24 OH or H$_2$CO MM sources in the sample.
Open symbols may also represent undetected OH MM because of
beaming effects or weak maser action with similar ISM conditions.
}
 \label{COlineratios}
\end{center}
\end{figure}

\begin{figure}[htbp]
\begin{center}
\includegraphics[width=7cm,clip]{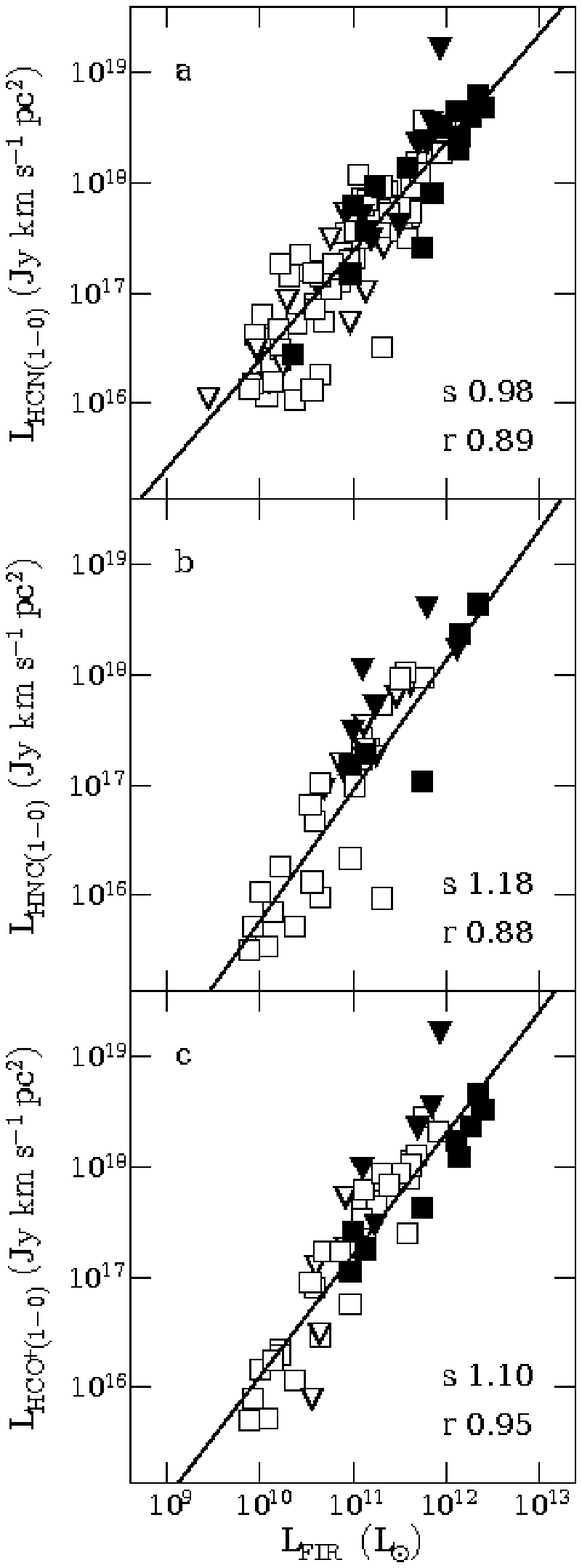}
\caption{{\bf a)} Integrated HCN\,(1$-$0), {\bf b)} integrated
HNC\,(1$-$0), and {\bf c)} integrated \HCOP\,(1$-$0) line
luminosity versus FIR luminosity for the total sample. There are
84 reliable data points for HCN, 27 for HNC, and 42 for \HCOP. The
slope of the fitted line ($s$) and the correlation coefficient
($r$) are shown in the lower right corner of each frame. The
displayed upper limits have not been used in the fits. The symbols
are explained with Figure 1.} \label{hd-vs-fir}
\end{center}
\end{figure}

\section{The CO line characteristics}
\label{sec:COlines}

\subsection{Line luminosities}

The $J$=1$-$0 and 2$-$1 mm-wave lines of CO observed by us are
easily saturated and thermalized, such that they are likely not
tracing the high density regions (n(H$_2$) $\geq$
10$^4$\,cm$^{-3}$) of the ISM where the star formation occurs. The
CO characteristics of our sample sources are shown in Fig.
\ref{COlineratios}.   Slopes (indicated in the figures with $s$)
have been fitted to the data using a standard Least Squares Fit
method \citep[see][]{Bevington1992}. To evaluate the significance
of the fits, the correlation coefficient (indicated by $r$) has
also been calculated.  For these calculations upper limits have
been omitted, and all data have been given the same statistical
weight, since only few data points have known errors. Both CO
transitions display a clear dependence on the FIR luminosity with
slopes of 0.74 $\pm$0.04 (110 sources; $r$ = 0.87) and 0.91 $\pm$
0.10 (32 sources; $r$ = 0.85).  At high luminosities the scatter
in $L_{\rm CO}$ is a factor 4 but increases at lower luminosities
to a factor of 10. This scatter can be attributed to variations of
the star formation efficiency (SFE) \irlums/$M$(H$_2$), ranging
between 1 and 100 \solums/\solmass, and (in part) to
underestimates of the CO luminosity for the lower luminosity (and
nearby) sources, where single-dish instruments only sample the
central region.

The slopes for both CO curves are significantly less steep than
unity, possibly related to optical depth effects or to an increase
in the star formation efficiency SFE at high \irlums. Assuming
constant $L_{\rm CO}$-to-$M$(H$_2$) conversions, earlier studies
suggested a linear \irlum - $M$(H$_2$) relation with a slope that
steepens at high \irlum  due to a strong variation of the SFE
\citep{YoungScoville1991}. Similarly \cite{BraineCombes1992}
present a linear relation between the CO(2$-$1) and FIR data for
\irlum $\leq$ 10$^{10.5}$ \solums. The relation with slope less
than unity deduced from our data for \irlum $\geq$ 10$^{10}$
\solum is consistent with characteristics of the upper luminosity
range of the earlier data. Our fit would suggest SFE $\propto$
\irlums$^{0.25}$ for the higher end of the \irlum distribution, or
an equivalent variation in the $L_{\rm CO}$-to-$M$(H$_2$)
conversion factor.

\subsection{The CO line ratio}
\label{sec:COratios}

The $I_{\rm CO}$(2$-$1)/$I_{\rm CO}$(1$-$0) line ratio has been
displayed for 32 sources of which five have values greater than 2
(OH MM IRAS\,15107+0724, NGC\,660, NGC\,3256, IRAS\,06259$-$4708,
and IRAS\,18293$-$3413; Fig. \ref{COlineratios}c). Earlier results
from nearby spiral galaxies \citep{RadfordDS1991,
BraineCombes1992} as well as the data from this study show a
concentration of data points up to 1.5 with a few outliers up to
three.

An observed line ratio of $\geq$ 1 for a homogeneous gas could
imply that the gas is optically thin, which is not realistic for
the nuclear environments considered in our sample. An optically
thick cloud model with external heating of small dense clouds,
i.e. with hot young stars blowing holes into the molecular
structures, can make the CO\,(2$-$1) line stronger than the
CO\,(1$-$0) line, if the temperature decreases by about 30$-$60\,K
(factor 2) from the $\tau$ = 1 surface for the CO\,(2$-$1) line to
the $\tau$ = 1 surface for the CO\,(1$-$0) line
\citep{BraineCombes1992}. The higher opacity of the CO\,(2$-$1)
line as compared with the CO\,(1$-$0) line, in almost all
scenarios, leads to a higher beam filling factor for the
CO\,(2$-$1) line and thus a line intensity ratio larger than
unity.

Early evaluation of the $I_{\rm CO}$ to hydrogen mass $M$(H$_2$)
conversion factor in galaxies by \cite{MaloneyB1988} assumed
thermalized CO rotational levels with $T_{\rm ex}$ = $T_{\rm
kin}$. This would predict CO\,(2$-$1)/CO\,(1$-$0) brightness
temperature ratios in the range of 0.9 to 1.1 in the Galaxy. In
order to explain similar extragalactic data, \cite{RadfordDS1991}
show that sub-thermal excitation ($T_{\rm ex}$ $<$ $T_{\rm kin}$)
in the $n$(H$_2$) = 10$^2$ to 10$^3$\,cm$^{-3}$ range could result
in integrated CO\,(2$-$1)/CO\,(1$-$0) line ratios in the 1.4$-$2
range, and beam-corrected (intrinsic) $T_{\rm b}$(2$-$1)/$T_{\rm
b}$(1$-$0) ratios in the range of 0.6$-$0.75.

An evaluation of the CO properties requires consideration of the
relative filling factor of the two CO transitions (Sec.
\ref{sec:results}). Here the conversion from the observed $T_{\rm
mb}$ ratio to an intrinsic $T_{\rm b}$ ratio involves an antenna
beam correction factor ($d_{\rm s}^2$ + $b_{\rm 21}^2$)/($d_{\rm
s}^2$ + $b_{\rm 10}^2$), which expresses the relative size of the
source $d_{\rm s}$ and the beam size $b$ at the two transitions.
For a representative size of 2 kpc for the prominent CO emission
regions in luminous FIR galaxies, the intrinsic $T_{\rm b}$ ratios
will be reduced by a factor of 1.15 - 2.46 (for velocities of 750
to 7500\,\kmss) relative to the corresponding $T_{\rm mb}$ ratios.
This range of values suggests that most of the observed ratios
could have intrinsic $T_{\rm b}$ ratios around unity, which is
consistent with a sub-thermal excitation of the low density
molecular medium \citep{RadfordDS1991}. Our most extreme value of
3.0 for the OH MM source IR\,15107+0724 at 46 Mpc would give a
brightness temperature ratio of 1.4.

The conversion of integrated CO\,(1$-$0) lines to H$_2$ column
densities is expressed as $N$(H$_2$)/$I_{\rm CO(1-0)}$ = $\alpha$
\citep{StrongEA1988}. These authors suggested a value $\alpha$ =
2.3 10$^{20}$ molecules cm$^{-2}$/(K km\,s$^{-1}$), using the
properties of self-gravitating virialized molecular clouds of high
optical depths. However, this value is not uniformly applicable
for the Galaxy and for extragalactic nuclei. Studies of the
Galactic Center using COBE data \citep{SodroskiEA1995} and
C$^{18}$O and the mm dust emission \citep{DahmenEA1998} find that
the conversion factor is a highly variable value, resulting in an
overestimate of the mass in the nuclear region and an
underestimate in the outer disk. Studies of the central regions of
the nearby galaxies NGC\,253 \citep{MauersbergerEA1996a} and
NGC\,4945 \citep{MauersbergerEA1996b} show an overestimate by a
factor of 6$-$10. Studies of ULIRGs show that most of the CO
luminosity comes from relatively low-density gas that is
dynamically significant and is located in a molecular disk
\citep{SolomonEA1997, DownesSolomon1998}. A conservative estimate
of the CO$-$to$-$H$_2$ conversion $M$(H$_2$)/$L_{CO}'$ = 0.8
\solmass /(K km s$^{-1}$ pc$^2$) for such extragalactic nuclei is
based on an $\alpha$ factor about 5 times lower than the value for
self-gravitating clouds. This conversion ratio may be used to
convert the CO\,(1$-$0) values listed in Table B.1 into H$_2$
masses.

\begin{figure}[htbp]
\begin{center}
\includegraphics[width=7cm,clip]{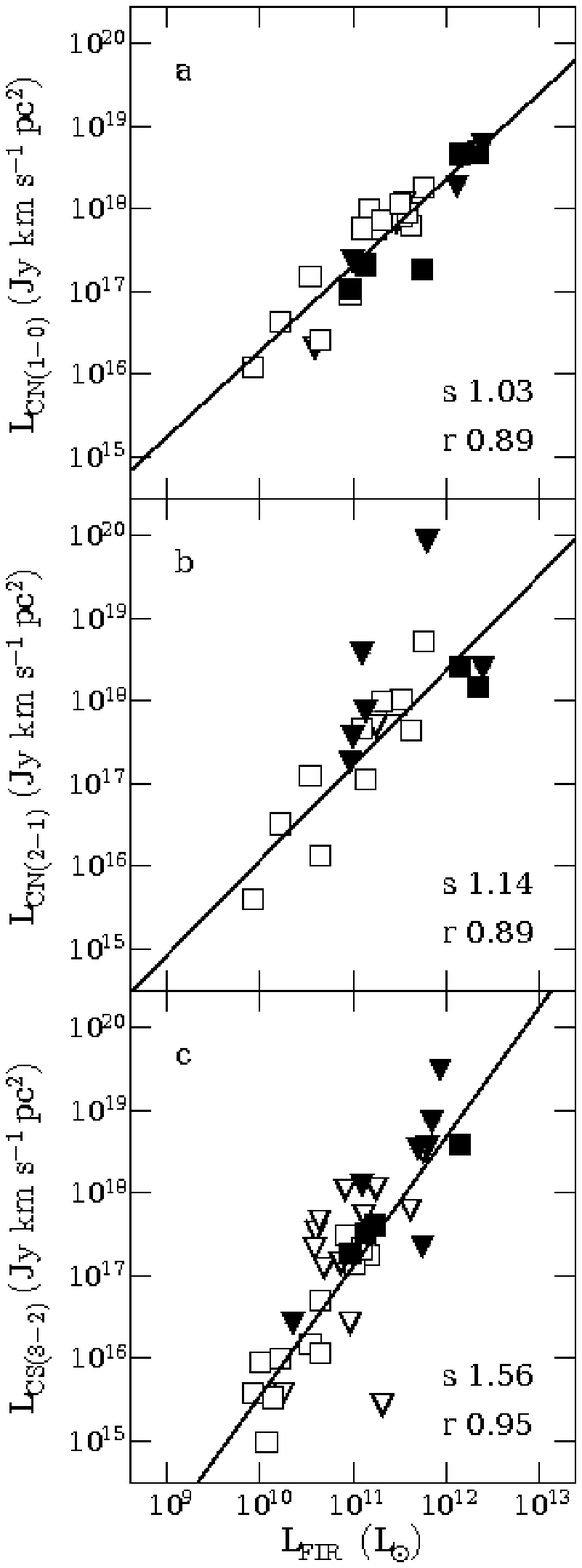}
\caption{{\bf a)} Integrated CN $N$ = (1$-$0), {\bf b)} integrated
CN $N$ = (2$-$1), and {\bf c)} integrated CS\,(3$-$2) line
luminosity versus FIR luminosity for the total sample. There are
19 reliable data points for CN(1$-$0), 13 for CN(2$-$1), and 17
for CS(3$-$2). The slope of the fitted curves ($s$) and the
correlation coefficient ($r$) are shown in the lower right corner
of each frame. No upper limits have been used in the fits.  The
symbols are described with Figure 1.} \label{cs&cn}
\end{center}
\end{figure}

\section{High-density tracer molecules}

\subsection{Line luminosities}

The diagrams of line luminosity against the infrared luminosity
for all molecular transitions considered (Figs.
\ref{COlineratios}, \ref{hd-vs-fir}, and \ref{cs&cn}) show a
well-behaved distribution with a considerable width. The spread of
the data points may be caused by a variety of effects, such as
non-radiative heating of the gas, differences in the initial
abundances, and the evolutionary changes of the ISM resulting from
a starburst and the depletion of the high-density material (see
Sec. \ref{sec:evolution} below). Data points at low luminosity
that lie well below the center of the line luminosity distribution
may result from using only the central pointing for nearby
galaxies.

{\bf The HCN, HNC and HCO$^+$ Molecules} - The line luminosities
of the source sample of the prominent tracer molecules are
presented against the FIR luminosity in Fig. \ref{hd-vs-fir}. The
slopes for these three molecules are close to unity such that HCN
(84 sources) has $s$ = 0.98 $\pm$ 0.06 ($r$ = 0.89), HNC (28
sources) has $s$ = 1.18 $\pm$ 0.13  ($r$ = 0.88), and \HCOP (42
sources) has $s$ = 1.10 $\pm$ 0.06 ($r$ = 0.95). The tracer
luminosities provide a slightly tighter relation with \irlum than
the CO(1$-$0) data as displayed in Fig. \ref{COlineratios}.  Fits
with slope 0.9 $\pm$ 0.1 for HCN versus \HCOP\ and with slope 1.4
$\pm$ 0.4 for HCN versus CO(1$-$0) have been noted by
\cite{NguyenEA1992}. A slope of 0.97 $\pm$ 0.05 was found for HCN
versus \irlum \citep{GaoS2004a} and slopes of 0.88 $\pm$ 0.15 for
HNC versus both HCN and \HCOP \citep{HuttemeisterEA1995}. The
scatter of the distributions is about one decade in the main body
of data points.

{\bf The CN Molecule} - The mm$-$wave spectra of the cyano radical
CN have more complex characteristics than those of the other
observed molecular species. The CN $N$=1$-$0 line observations at
Pico Veleta have been taken at the strongest feature, the
$J$=3/2$-$1/2 transition at 113.491 GHz that contains one third of
the total intensity in the optically thin limit under Local
Thermodynamic Equilibrium (LTE) conditions. For the SEST data for
NGC\,1068, besides the $J$=3/2$-$1/2 feature a second
$J$=1/2$-$1/2 feature is seen at 113.191 GHz at a relative
strength of 0.45 (Figs. A.2). This relative strength is close to
the LTE value, which could suggest that the lines are optically
thin \citep[see][]{WangEA2004}. The CN $N$=2$-$1 transition data
focus on the $J$=5/2$-$3/2 line at 226.659\,GHz that represents
one sixth of the total intensity ($\tau <$1; LTE). The CN data has
been augmented with data for both transitions from
\cite{AaltoPHC2002}.

Most of the CN detections are single broad features centered at
the systemic velocity of the galaxies (Figs. A.1$-$2). Only Arp
\,220 and IR\,05414 show double $N$ = 1$-$0 and $N$ = 2$-$1 CN
components that cannot be attributed to other spingroups but only
to the two nuclei of their respective systems.

The CN line luminosities in Figure \ref{cs&cn}ab show a
distribution that is similar to that of other tracer molecules. A
slope close to unity is found for CN(1$-$0) (19 sources): $s$ =
1.03 $\pm$ 0.13 ($r$ = 0.89) and for (CN(2$-$1) (13 sources): $s$
= 1.14 $\pm$ 0.17 ($r$ = 0.89). The scatter of the distributions
is less than one order of magnitude.

{\bf The CS Molecule} - Our CS\,(3$-$2) data from the PV 30-m and
SEST telescopes (Figs. A.1$-$2) have been augmented with
detections from \cite{MauersbergerHWH1989},
\cite{MauersbergerH1989}, and \cite{WangEA2004}. An early
assessment of the CS emission in nearby galaxies shows a clear
relation with \colum\, and \irlum and the star formation rate
\citep{MauersbergerH1989}. The CS line luminosity (17 sources)
versus the FIR luminosity has a higher slope, $s$ = 1.56 $\pm$
0.14 ($r$ = 0.95), as compared with a slope of unity for the much
smaller sample of \cite{MauersbergerHWH1989}. This larger slope of
CS as compared to the other molecules under consideration
indicates a stronger environmental dependence (Fig. \ref{cs&cn}c).

\begin{figure}[htbp]
\begin{center}
\includegraphics[width=7cm,clip]{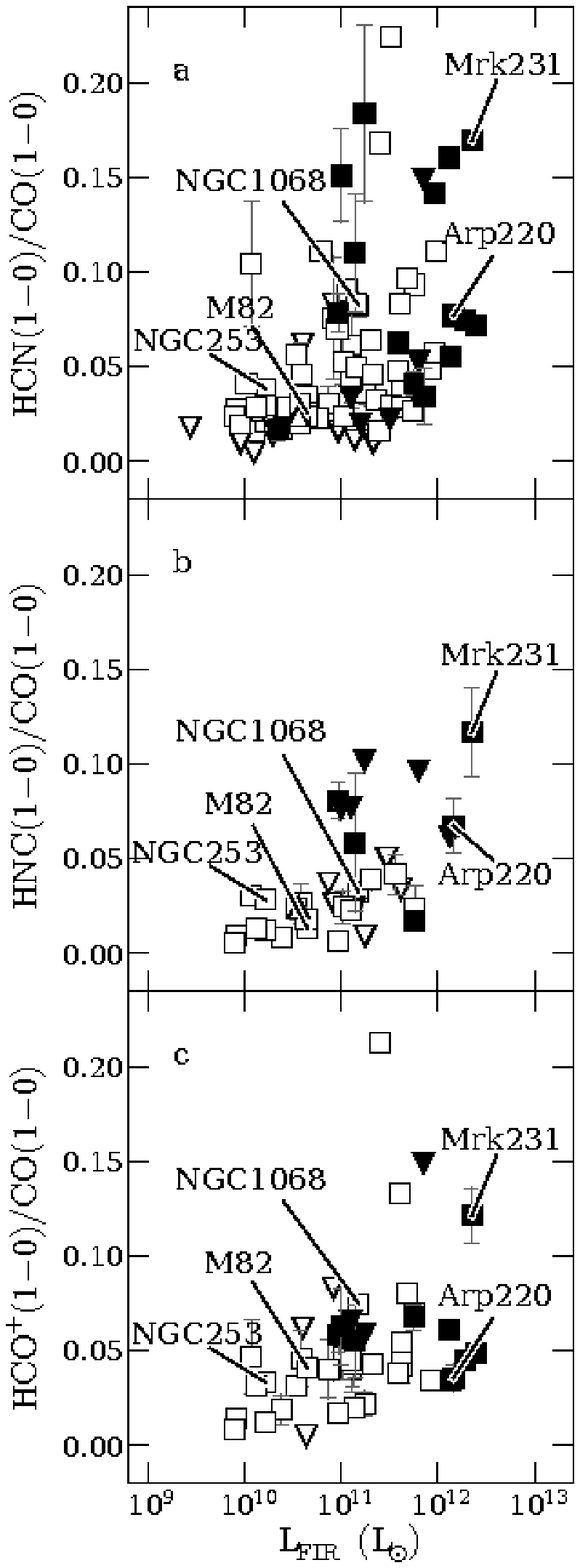}
\caption{The fraction of the high-density component relative to
the low-density component. {\bf a)} Integrated
HCN\,(1$-$0)/CO\,(1$-$0), {\bf b)} integrated
HNC\,(1$-$0)/CO\,(1$-$0), and {\bf c)} integrated
\HCOP\,(1$-$0)/CO\,(1$-$0) line ratios as a function of FIR
luminosity. The symbols are described with Figure 1.}
\label{hd_co-vs-fir}
\end{center}
\end{figure}

\subsection{Evaluating the relations}

The well-known Kennicutt-Schmidt laws relate the {\it
surface-density of star formation} ($\Sigma_{\rm SFR}$) and the
{\it surface-density of the gas} ($\Sigma_{gas}$). On the basis of
H\,I, CO, and H$\alpha$ data for a large sample of spiral and
starburst galaxies, \cite{Kennicutt1998} finds that the
$\Sigma_{\rm SFR}-\Sigma_{gas}$ relation has an exponent
$\alpha$=1.4 ($\Sigma_{\rm SFR}\propto\Sigma_{gas}^{\alpha}$).
Analogously the relation of the integrated luminosity of the
high-density gas component involved in the star formation process
and the integrated star formation rate (SFR) expressed as \irlum
could also behave as a Kennicutt-Schmidt law.
\cite{KrumholzThompson2007} made model predictions for the
Kennicutt-Schmidt laws assuming a constant star formation
efficiency.  They find that the line luminosity for molecules with
a critical density lower than the average density of the system
rises linearly with density. For lines with critical densities
larger than the median gas density the luminosity per unit volume
increases faster than linearly with the density.  The star
formation rate also rises super-linearly with the gas density, and
the combination of these two effects produces a close to linear
correlation ($\alpha\sim1$) for the high critical density
molecules and a super-linear relation for the low density tracers
($\alpha \sim1.5$; the classic Kennicutt-Schmidt law).  They also
predict that generally there may be more intrinsic scatter in this
relation for low critical density molecules than for high density
tracers.

Our data mostly confirms the findings of
\cite{KrumholzThompson2007}. Our CO distributions (Fig.
\ref{COlineratios}) show indices of $\alpha$=1.40 and 1.23 (the
inverse of the slope $s$ in our diagrams) and have a relatively
large spread. The relations between \irlum and the line
luminosities of high density tracers HCN\,(1$-$0), HNC\,(1$-$0),
\HCOP\,(1$-$0), CN\,(1$-$0), and CN\,(2$-$1) are all linear within
the errors (see Figs. \ref{hd-vs-fir} and \ref{cs&cn}).  However,
the relation for CS\,(3$-$2) is sub-linear with $\alpha$=0.66.
Since CS has a critical density close to that of HCN and higher
than that for CO, CS is expected to have a linear relation
\citep{KrumholzThompson2007}. This may indicate that this species
is influenced by additional chemical and environmental effects.

\begin{figure}[htbp]
\begin{center}
\includegraphics[width=7cm,clip]{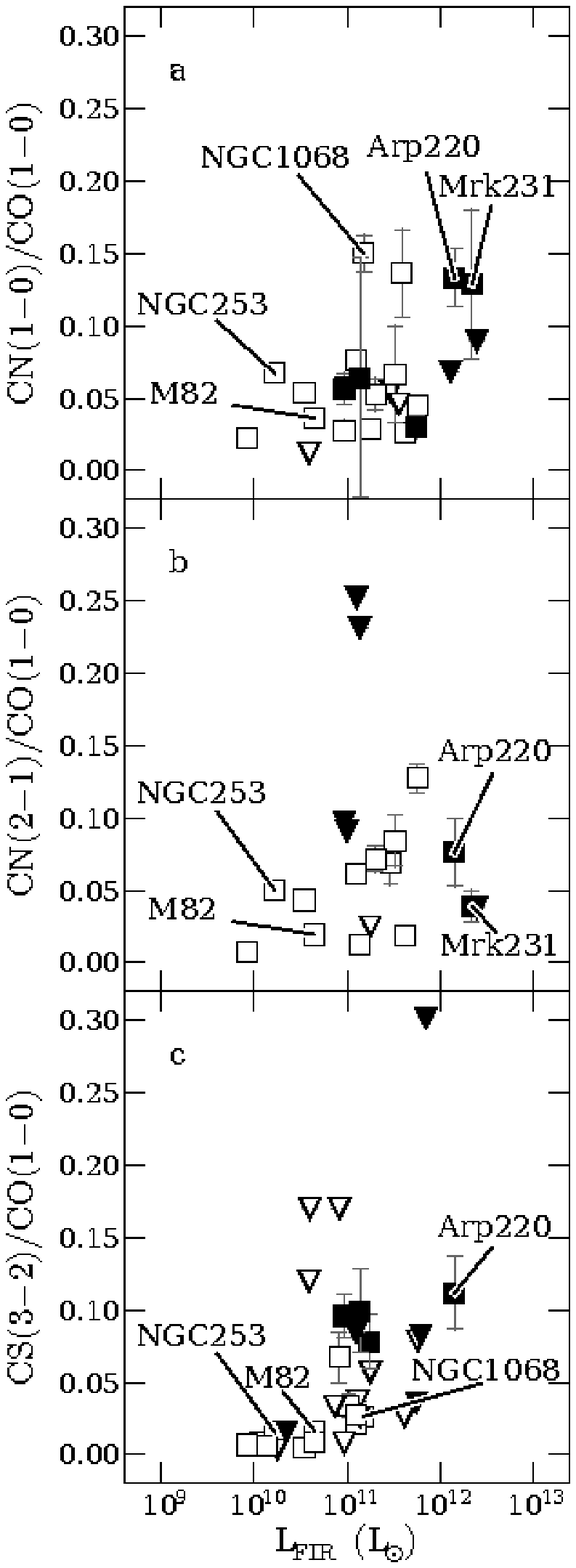}
\caption{The fraction of the high-density component relative to
the low-density component. {\bf a)} Integrated
CN\,(1$-$0)/CO\,(1$-$0), {\bf b)} integrated
CN\,(2$-$1)/CO\,(1$-$0), and {\bf c)} integrated
CS\,(3$-$2)/CO\,(1$-$0) line ratios versus FIR luminosity. The
symbols are described with Figure 1.} \label{FIR_cncs_co}
\end{center}
\end{figure}

\subsection{Tracer intensities relative to CO\,(1$-$0)}
\label{sec:HD/COevol}

In Fig. \ref{hd_co-vs-fir} and \ref{FIR_cncs_co} the integrated
intensities for the total sample are compared with the CO(1$-$0)
emission that represents the larger scale low-density component.
Ignoring NGC\,7469, which seems to have unusual high ratios, and
upper limits, the emission line ratio varies from 0.01 to 0.18 for
HCN, from 0.01 to 0.08 for HNC, from 0.01 to 0.21 for \HCOP, from
0.01 to 0.15 for CN, and from 0.01 to 0.11 for CS. An early study
of CS suggests a relative line ratio in the 0.01$-$0.08 range
\citep{MauersbergerH1989}. The upper range for the HCN/CO
intensity ratio of our current sample is comparable with the
sample of \cite{GaoS2004a}. The range for HNC and \HCOP\ is found
to be (slightly) smaller than for HCN (when ignoring the
outliers/upper limits) \citep[see also][]{GraciaCarpioGPC2006}.

We note that the distribution of characteristic line ratios for
all molecules increases with FIR luminosity, which gives an
upwardly curved lower boundary for the distribution at higher
\irlums. The highest values for all ratios are found at \irlum
$\geq 10^{10.5}$ \lsols. {As discussed in Sec. \ref{sec:evolution}
below, a decreasing line strength of a high-density tracers as
compared with CO during the course of a nuclear starburst would be
the signature of the depletion of the nuclear high-density
component.}

\section{Modelling of a ULIRG outburst}
\label{sec:evolution}

\subsection{The outburst model}

The FIR light curve of an ULIRG is determined by the radiative
energy generated by a (circum-)nuclear starburst possible in
hybrid mode with an AGN, that is re-radiated by a torus-like dusty
environment. During such an FIR outburst, the low-density and
high-density molecular components of the nuclear region evolve
differently. The large-scale signature of the low-density gas, as
represented by CO\,(1$-$0) emission, would only be partially
altered by the centralized star formation process. On the other
hand, the centrally concentrated high-density gas forms the
structural basis for the star formation process and will be slowly
consumed, destroyed or dispersed by the newly formed stars.

In a simplified scenario, the instantaneous FIR luminosity of the
ULIRG represents the rate at which the nuclear activity generates
FIR energy and consumes/destroys its high-density molecular
material. In the absence of a representative FIR light curve with
different timescales for the rise and fall of the FIR luminosity,
we choose a diffusion-like expression as a response to a sudden
start of the star formation. This function may be defined for $t$
$>$ 0 as:

\begin{equation} L_{\rm FIR}(t) = 1.35\,\,L_{\rm FIR}(max)
\left(\frac{T}{t}\right)^{2.5} e^{-T/t}, \label{eq:1}
\end{equation}

\noindent where $L_{\rm FIR}$(max) is the maximum luminosity and
$T$ is the characteristic timescale of the burst. The duration of
the burst (down to 1\% of the peak) is on the order of 10 $\times$
$T$. A diffusion curve may not be the most appropriate
representation of a FIR light curve but it resembles those
obtained with starburst-driven models of the FIR evolution and
color properties of ULIRGs \citep{LoenenBS2006}.

\begin{figure}[htbp!]
\begin{center}
\includegraphics[width=6.5cm,clip]{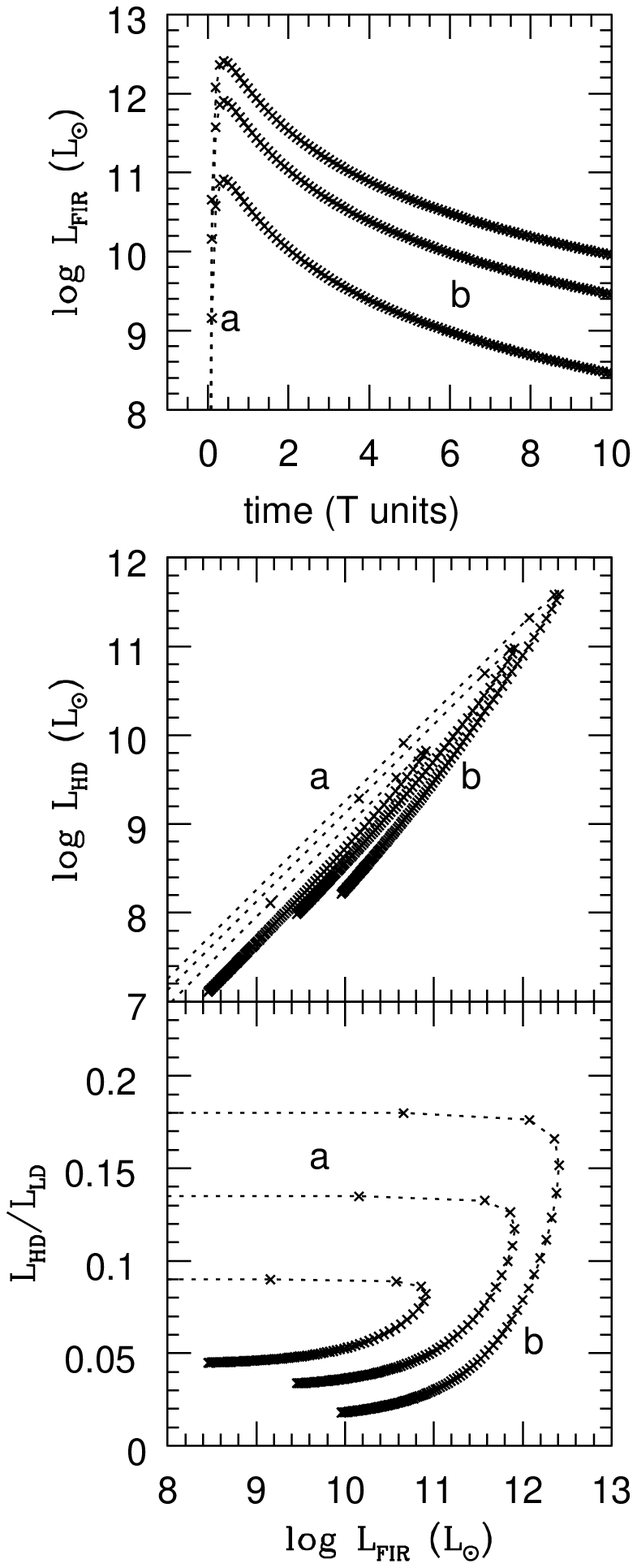}
\caption{Simulations of characteristic behavior of the
high-density component during an FIR outburst. Labels (a) and (b)
indicate the rapidly rising and the slower decaying segments of
the outburst, respectively. {\it Upper diagram}: Three \irlum
light curves versus time on a log-linear scale. The time is in
units of $T$ as used in Eq. 1. {\it Middle diagram}: The
associated luminosity of the high-density ($HD$) component $L_{\rm
HD}$ for the three light curves on a log-log scale. The parameters
of the curves as described in Sec. \ref{sec:evolution} are: 1) the
outer curve with initial $\beta_0$ = 0.18, log \irlums(max) =
12.5, and a depletion rate of $\Gamma$ = 0.9 during the course of
the outburst; 2) the middle curve 0.135, 12.0, and 0.5; and 3) the
inner curve 0.09, 11.0, and 0.4. {\it Lower diagram}: The
time-evolution of the ratio of high- and low-density (LD)
components for the three FIR light curves of the above frames. }
\label{simul1}
\end{center}
\end{figure}

The initial line strength of the high-density $HD(t=0)$ component
can be defined as a fraction $\beta_{\rm 0}$ of that of $LD$, the
larger-scale low-density component, that is assumed to remain
unchanged during the outburst. As a result, the line ratio of the
high- and low-density components varies with time during the FIR
outburst as:

\begin{equation} \frac{HD(t)}{LD} = \beta_{\rm 0} \left[1 - \Gamma
\frac{\int^t_0{L_{\rm FIR}(t) {\rm d}t}}{E_{\rm FIR,int}}\right],
\label{eq:2}
\end{equation}

\noindent where $\Gamma$ is the total fraction of the initial
$\beta_{\rm 0}$ of the $HD$ emission component that is destroyed
(disappears) during the whole course of the outburst. The
instantaneous $HD$ component depends on the fraction of the FIR
energy generated up till time $t$ and the total FIR energy
radiated during the outburst $E_{\rm FIR,int}$.

\subsection{Modelling results}

The results obtained in these simulations are presented in Fig.
\ref{simul1} for three FIR light curves with peak luminosities log
$L_{\rm FIR}(max)$ = 11, 12, and 12.5 in order to cover the range
of observed FIR luminosities. The FIR light curves (upper frame)
are used to determine the evolving luminosity of a high-density
component (middle frame), and the evolving emission ratio of the
high- and low-density components (bottom frame). A light curve
having a more delayed peak would give more rounded curves and the
turn-around point would occur at higher $HD(t)/LD$ values.

This simple time-dependent scenario already shows some distinctive
behavior of the high-density component during a FIR outburst, that
is relevant to the high-density tracer data presented in this
paper:
\newline \noindent {\bf 1)} The {\it rapid rise segment}
(indicated as (a) in Fig. \ref{simul1}) of the star formation
process and the resulting FIR outburst ensure that the data points
show rapid evolution and are relatively sparse at the onset of the
curves. During the peak of the curve the depletion rate of the
high-density component is highest. After the luminosity peak the
data points in the {\it decay segment} of the three curves
(indicated as (b) in Fig. \ref{simul1}) will be closer together.
Because of the increased probability of finding a source in this
state, our source samples will be dominated by sources in the
intermediate and later stages of evolution.
\newline \noindent {\bf 2)} The luminosity of the high-density
component versus \irlum will display a separation between the
upward (a) and downward (b) luminosity segment due to the ongoing
depletion (middle frame Fig. \ref{simul1}). Sources with the
highest peak luminosities and the largest depletion factors follow
the outermost evolutionary track and will end on the lower
boundary of the distribution. This effect could add significantly
to the scatter in the diagrams of the high density tracers versus
\irlums.
\newline \noindent {\bf 3)} The high- to low-density emission ratio
versus FIR (bottom frame Fig. \ref{simul1}) reflects the phases of
the FIR light curve: a rapid horizontal track for the {\it rapid
rise segment (a)} of the FIR luminosity with few data points,
followed by a downward curved track with accumulation of data
points at the lower boundary for the {\it decay segments (b)}.
After finishing the FIR light curve, a third {\it buildup phase}
may take place with a steady re-building of the high-density
molecular component in the central regions. A molecular buildup
would prepare the galaxy for the next star formation outburst.

\subsection{Comparison with observations}

The observed relations between molecular line luminosities and FIR
luminosities in Figs \ref{COlineratios}, \ref{hd-vs-fir} and
\ref{cs&cn} are in agreement with the simulations in the middle
frame of Fig. \ref{simul1}. The observed scatter in the diagrams
would partly result from differences in the rise and decay stages
of an outburst. Furthermore, the track of the highest luminosity
sources and OH MM would lie at the outer edges of the
distribution. Similarly, the distribution of data points in an
$L_{HD}$ - $L_{\rm FIR}$ diagram could display a steepening
towards higher \irlum values (see Fig. \ref{hd-vs-fir}).

The observed intensity ratios of tracer molecules relative to
CO\,(1$-$0) show organized distributions (Figs. \ref{hd_co-vs-fir}
and \ref{FIR_cncs_co}), that are in agreement with the simulations
in Fig. \ref{simul1}c. The upper regions of the distributions are
indeed sparsely populated as they represent (near-)initial values
for sources. From the diagrams it follows that the initial
fraction $\beta_0$ of the high-density component would be in the
0.08$-$0.18 range, and the depletion rate $\Gamma$ for our sample
of sources would lie in the 0.6$-$0.9 range. Further evolution of
the outburst during the decay phase would result in the
characteristically curved lower boundary observed in the diagrams.
The highest FIR luminosity sources that have passed beyond their
peak are also found at this lower boundary.

The distribution of OH MM (and H$_2$CO MM) are quite conspicuous
in the diagrams. Some occupy the sparse region of rapid evolution
at the highest $HD$/$LD$ intensity ratios (high $\beta_{\rm 0}$),
while others have already evolved to the curved lower edge of the
distribution, as seen clearly for the HCN/CO ratio in Fig.
\ref{hd_co-vs-fir}a.

\section{Molecular excitation environments}
\label{sec:pumping}

The emission properties of Galactic and also extragalactic
emission regions have been studied and interpreted by using two
distinct excitation regimes: X-ray Dominated Regions (XDRs) and
Photon Dominated Regions (PDRs) with far-ultraviolet radiation
fields. These environments are quite distinguishable using the
emission characteristics of high-density tracers in the cloud core
regions and low-density tracers at the cloud surfaces
\citep{1996A&A...306L..21L, MeijerinkS2005, BogerS2005}. The
abundance and emission characteristics of the molecular and atomic
lines depend strongly on the available (attainable) column density
submitted to the radiation field and the strength of the radiation
field.

{\bf Global environmental conditions} - The PDR and XDR modelling
of \cite{MeijerinkS2005} and \cite{MeijerinkSI2006,
MeijerinkSI2007} incorporate the physics and the chemistry in the
local environment using many molecules and chemical reactions
under local thermal balance. A large grid of density and radiative
conditions was used to predict the cumulative column densities
taking into account the radiation transfer in a representative
slab-model incorporating XDR and PDR environments. This grid
predicts column densities in all relevant energy states for the
dominant molecules. All high density tracer emissions are
predominantly produced by regions with high hydrogen column
densities. Relying on the predicted column densities for such a
gas component, we restrict ourselves to a first order analysis of
the observed line ratios using the ratios of calculated column
densities.

The characteristics of both the PDR and XDR models depend on the
parameterization of the radiative energy deposition rate per
hydrogen nucleus $P$ = $F_{\rm FUV}$/$n_{\rm H}$ or $F_{\rm
X}$/$n_{\rm H}$ and suggest the following regimes:

{\bf 1)} For PDRs with high values of $P$ $\geq$ 5 $\times$
10$^{-4}$ erg cm s$^{-1}$ the column density ratios vary for all
column densities $N_{\rm H}$ $\geq$ 10$^{\, 21.5}$ cm$^{-2}$ as:

 \vspace{0.2cm}
 \indent  $N$(\HCOP)/$N$(HCN) $<$  $N$(HNC)/$N$(HCN) $\leq$ 1
 \vspace{0.2cm}

\noindent while for smaller values for $N_{\rm H}$, the ratios
switch around such that:

\vspace{0.2cm}
 \indent   $N$(HNC)/$N$(HCN)  $<<$ 1 $\leq$  $N$(\HCOP)/$N$(HCN)
 \vspace{0.2cm}

\noindent and for lower $P$ $\approx$ 5 $\times$ 10$^{-6}$ erg cm
s$^{-1}$ the column density ratios vary as:

\vspace{0.2cm}
 \indent   1 $\leq$ $N$(HNC)/$N$(HCN)  $\approx$  $N$(\HCOP)/$N$(HCN)
 \vspace{0.2cm}

{\bf 2)} For XDRs the radiation penetrates deeper into the cloud
volume than for PDRs. For stronger XDR radiation fields ($P$
$\geq$ 5 $\times$ 10$^{-4}$ erg cm s$^{-1}$) and for column
densities above the transition point $N_{\rm H}$ $\approx$
10$^{22.5}$ cm$^{-2}$, the observable molecular column densities
have ratios:

 \vspace{0.2cm}
 \indent 1 $\leq$ $N$(HNC)/$N$(HCN) $\leq$ $N$(\HCOP)/$N$(HCN),
 \vspace{0.2cm}

 \noindent and below this $N_{\rm H}$ transition point:

 \vspace{0.2cm}
 \indent $N$(HNC)/$N$(HCN) $<<$ $N$(\HCOP)/$N$(HCN) $\leq$ 1,
 \vspace{0.2cm}

\noindent while for lower $P$ $\approx$ 5 $\times$ 10$^{-6}$ erg
cm s$^{-1}$ the column density ratios vary as:

\vspace{0.2cm}
 \indent   $N$(\HCOP)/$N$(HCN) $<<$ $N$(HNC)/$N$(HCN)  $\approx$  1
 \vspace{0.2cm}

{\bf 3)} For PDR conditions the column density ratio
$N$(CN)/$N$(HCN) varies from 0.5 to 2.0 over the respective
density range of 10$^{6}$ to 10$^{4}$ cm$^{-3}$.

{\bf 4)} For XDR conditions the column density ratio is
$N$(CN)/$N$(HCN) $>>$ 1 and may become as large as 10.

{\bf 5)} For PDR conditions at $N_{\rm H}$ $\geq$ 10$^{22}$
cm$^{-2}$ the column density ratio $N$(CS)/$N$(HCN) $\geq$ 1. At
lower column densities the ratio may become $\leq$ 1, which makes
CS a column density indicator.

\begin{figure}[htp!]
\begin{center}
\includegraphics[width=7cm,clip]{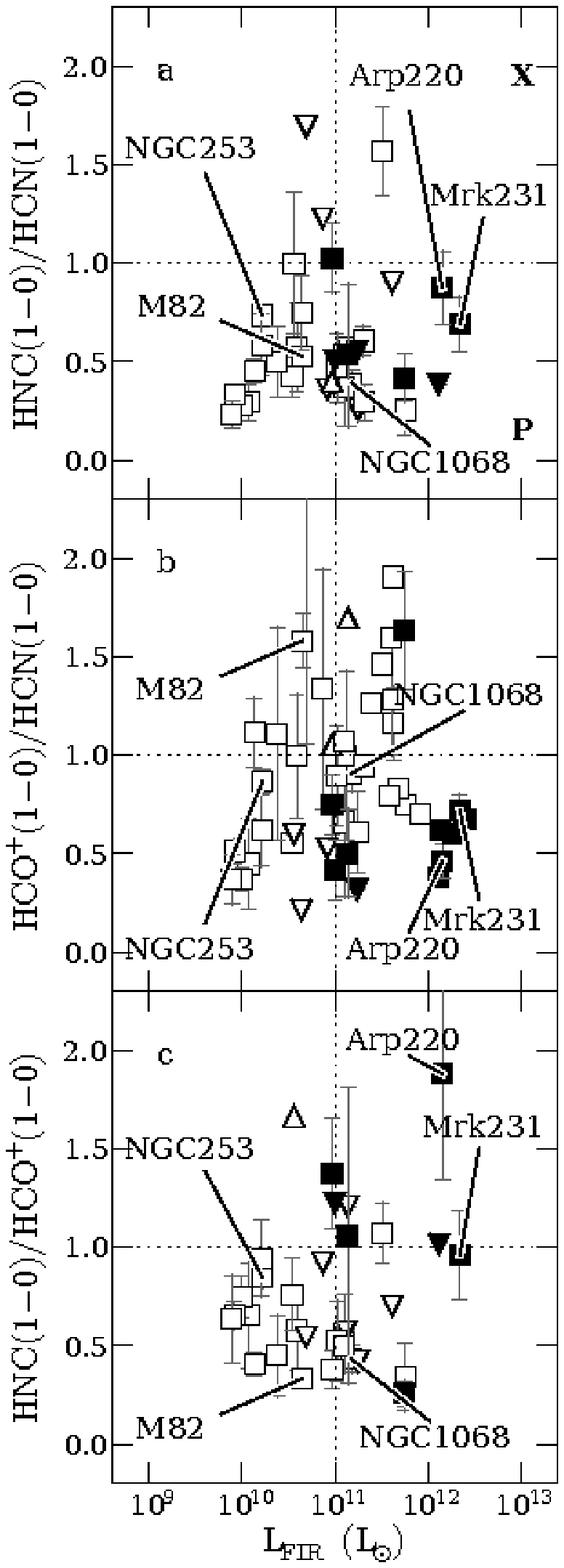}
\caption{High-density tracer line ratios versus FIR luminosity.
{\bf a)} the integrated HNC\,(1$-$0)/HCN\,(1$-$0) line intensity
ratio, {\bf b)} the integrated \HCOP(1$-$0)/HCN\,(1$-$0) line
intensity ratio,  and {\bf c)} the integrated
HNC\,(1$-$0)/\HCOP(1$-$0) line intensity ratio. The dotted lines
mark the predicted range of XDR-PDR excitation using {\it first
order} modelling diagnostics. In the upper frame the PDRs lie
below the line and the XDRs above. These labels have been omitted
in the two lower frames (where X would be in the upper range of
frame b) and in lower range of frame c)) because non-standard
heating effects on \HCOP and HNC (see section \ref{sec:HD-irlum})
cause a shift of the data points and makes labelling
non-appropriate. The symbols are described with Figure 1.}
\label{FIR_hd_hd}
\end{center}
\end{figure}

{\bf 6)} For XDR conditions for the whole column density range at
low values of $P$ $\approx$ 5 $\times$ 10$^{-6}$ erg cm s$^{-1}$
the column density ratio is $N$(CS)/$N$(HCN) $\geq$ 1. At higher
values of $P$ the ratio remains $\geq$ 1 for $N_{\rm H}$ $\geq$
10$^{23}$ cm$^{-2}$ and is $\leq$ 1 for lower $N_{\rm H}$, again
making it column density sensitive.

\begin{figure}[htp!]
\begin{center}
\includegraphics[width=7cm,clip]{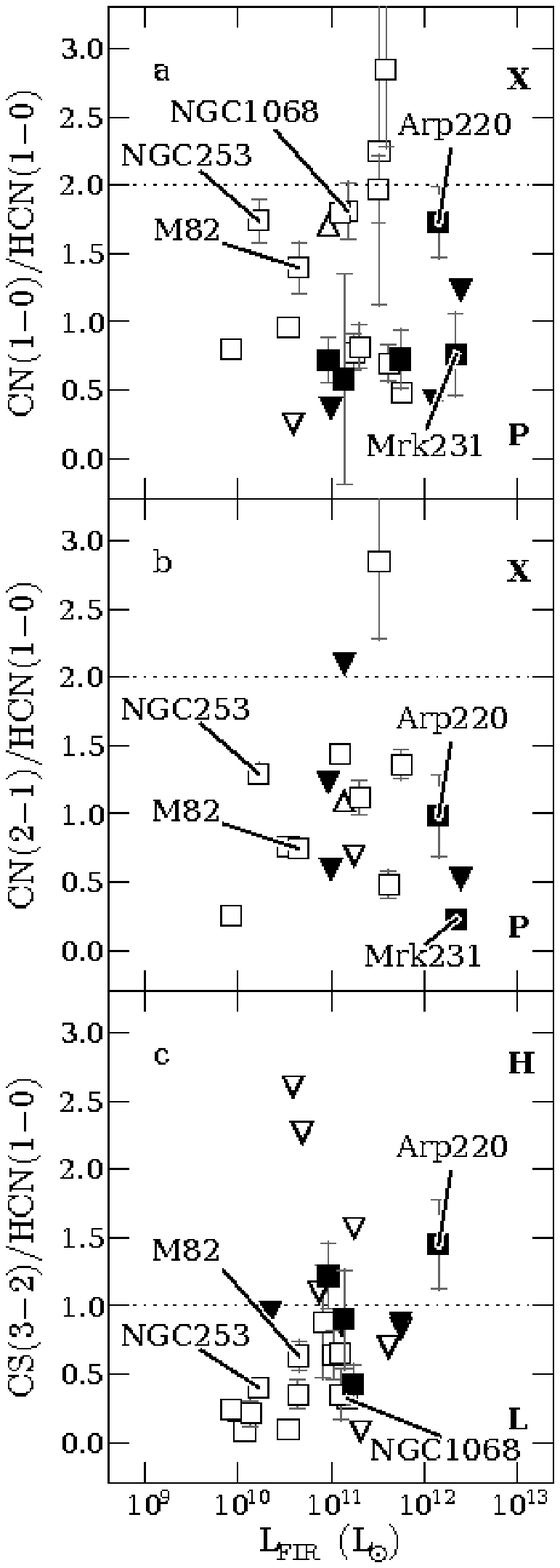}
\caption{CN/HCN and CS/HCN line ratios versus FIR luminosity. {\bf
a)} The integrated CN\,(1$-$0)/HCN\,(1$-$0) line intensity ratio,
{\bf b)} the integrated CN\,(2$-$1)/HCN\,(1$-$0) line intensity
ratio, and {\bf c)} the integrated CS\,(3$-$2)/HCN\,(1$-$0) line
intensity ratio. The symbols X and P and the dotted lines mark the
range of values for XDRs and PDRs. For the CS/HCN diagram the
dotted line divides the range for low- and high-column densities
(L \& H). The symbols are described with Figure 1.}
\label{FIR_cncs_hcn}
\end{center}
\end{figure}

{\bf Evolutionary environmental conditions} - The characteristics
described above are representative of a steady state nuclear
excitation environment. Realistically the behavior of molecules
tracing the high-density component is affected by changing
evolutionary and environmental effects that influence their
abundance and the exciting radiation fields.

Assuming that FIR luminous sources are mostly powered by
(circum-)nuclear starbursts, the evolving star formation will
generate supernovae with shocks enhancing the cosmic ray flux, and
produce new generations of X-ray binaries. Because these star
formation products alter the excitation environment, an
environment initially dominated by PDR conditions could change
into one with a combination of PDR and XDR conditions during the
later stages of the outburst \citep[for M\,82
see:][]{GarciaBEA2002,SpaansM2007}. This will modify the line
ratios from their steady state conditions discussed above. For
instance, increased cosmic ray production in shocks would enhance
the relative \HCOP abundance while shocks also would decrease the
relative HNC abundance \citep{Wootten1981, Elitzur1983,
SchilkeEA1992}. We will refer to this evolved starburst stage as
being in a ''late-stage'' as compared with an ''early-stage''
starburst where these effect do not yet play a role.

During the late-stage starburst also the density and temperature
structure of the ISM and its clumping would be affected, resulting
in chemical changes such as the selective destruction of HNC in
favor of HCN at higher temperatures (see Sec.
\ref{sec:HCNexcitation}). In addition, a certain fraction of the
high-density medium will be used/destroyed by the star formation
process. As a result the line emission may originate in regions of
decreasing column densities, which modifies the observed line
ratios.

\section{Diagnostics with tracer lines}

In the absence of multi-transition studies using LVG simulations
involving higher $J$ transitions, we make the assumption that the
integrated line ratios have a diagnostic value that is equivalent
to the ratios of column densities for the observable molecular
species.  The global characteristics of molecular column densities
described in the previous section for XDR- and PDR-dominance may
thus be used for a first diagnosis of the excitation environments.

\begin{figure*}[htbp]
\begin{center}
\includegraphics[width=12cm,clip]{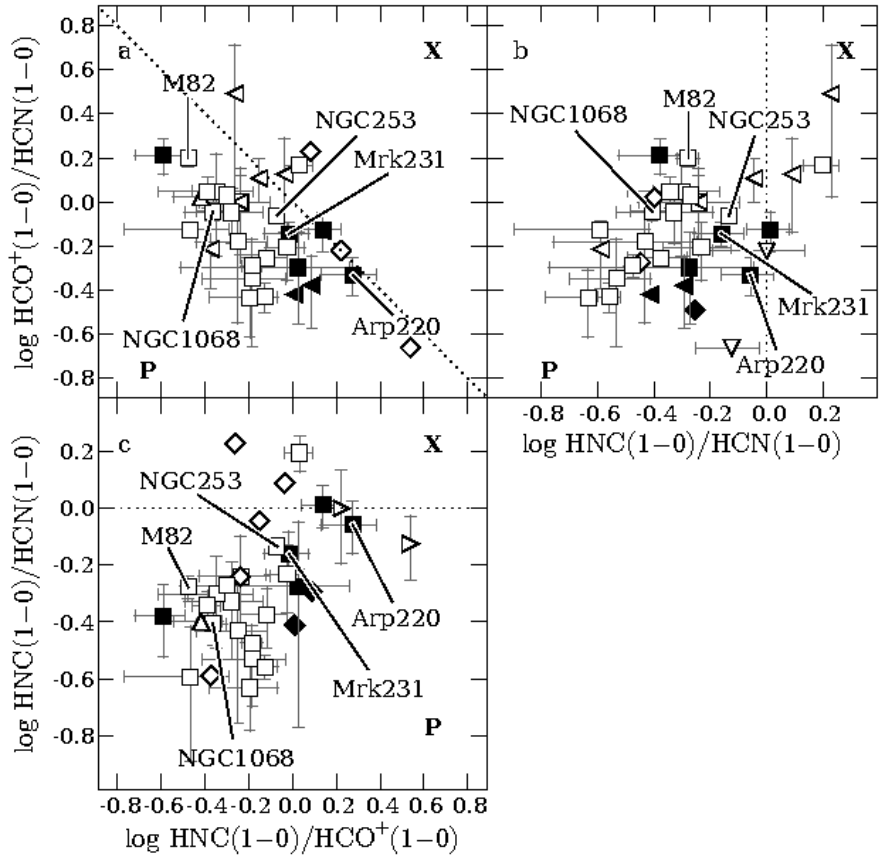}
\caption{The integrated line ratios of HCN\,(1$-$0), HNC\,(1$-$0),
and \HCOP\,(1$-$0) versus each other. {\bf a)} Integrated
\HCOP\,(1$-$0)/HCN\,(1$-$0) versus HNC\,(1$-$0)/\HCOP\,(1$-$0)\
ratios, {\bf b)} Integrated \HCOP\,(1$-$0)/HCN\,(1$-$0) versus
HNC\,(1$-$0)/HCN\,(1$-$0) ratios, and {\bf c)} Integrated
HNC(1$-$0)/HCN\,(1$-$0) versus HNC\,(1$-$0)/\HCOP\,(1$-$0) ratios.
The P and X and the dotted lines mark the regions of the line
ratios of predominantly photon-dominated regions (PDRs) and
predominantly X-ray-dominated regions (XDRs) using only the
HNC/HCN ratio. The symbols are described with Figure 1.}
\label{diagnostic1}
\end{center}
\end{figure*}

\subsection{High-density tracer ratios versus \irlum}
\label{sec:HD-irlum}

 Diagrams of relative intensities for HCN, HNC,
\HCOP, CN, and CS are presented in Figs. \ref{FIR_hd_hd} and
\ref{FIR_cncs_hcn}. Horizontal lines for a ratio equal to unity
(and two for CN) have been added to the diagrams in order to test
a {\it first order division} between PDR and XDR line ratios as
derived in the previous section. However, we find that the unity
value is not a reliable boundary for ratios with \HCOP. For CS
this line separates high and low column density. The line ratios
are predominantly determined by regions having the highest
molecular column densities \citep{MeijerinkSI2007}. All diagrams
show some systematic behavior in that the OH MM sources cluster
together and that there could be a wedge-like upper boundary for
the sources at high luminosities.

\begin{figure*}[htbp]
\begin{center}
\includegraphics[width=7.7cm]{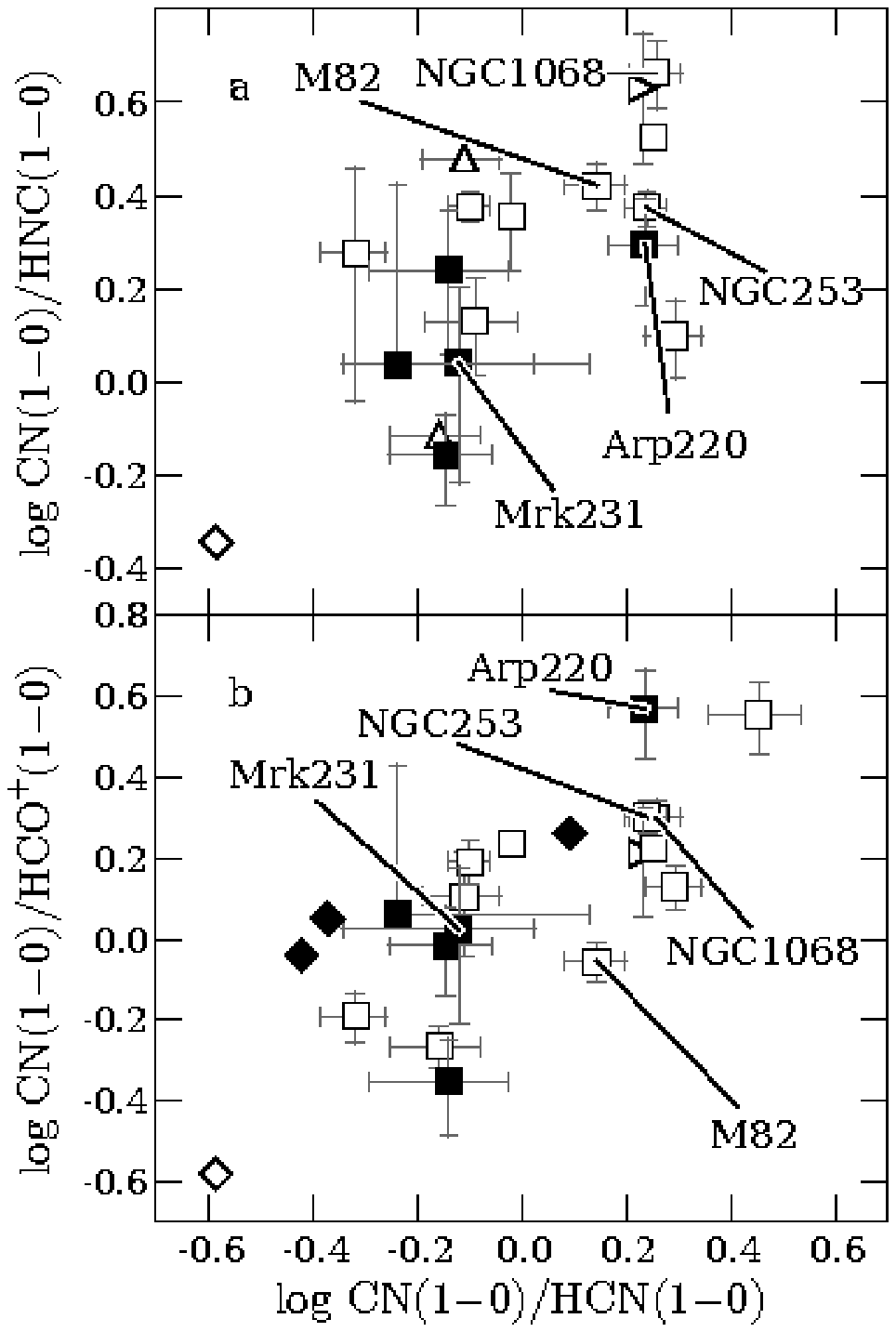}
\includegraphics[width=7.7cm]{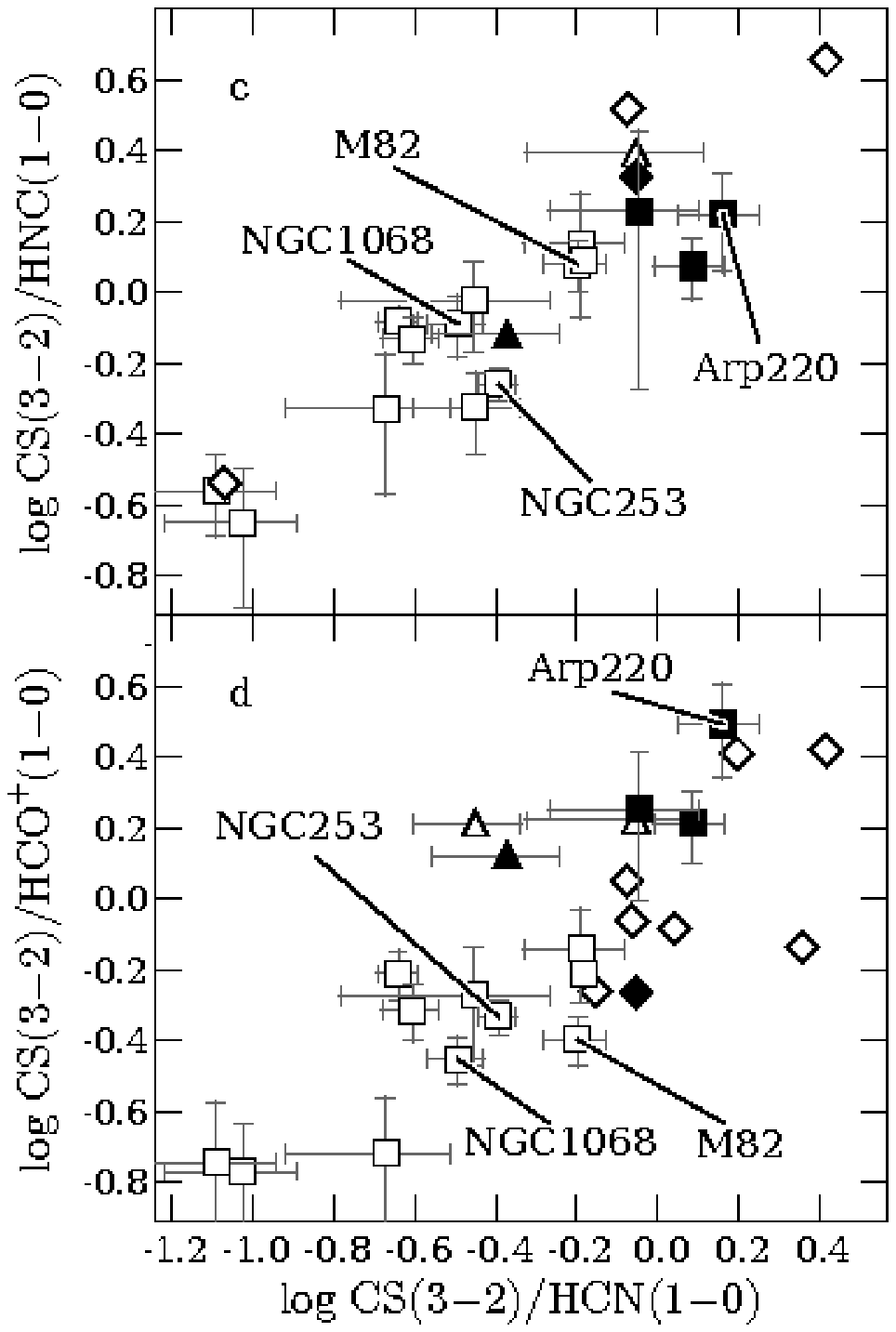}
\caption{The CN\,(1$-$0) and CS\,(3$-$2) lines compared with other
high-density tracers. {\bf a)} The CN/HCN ratio versus the CN/HNC
ratio and {\bf b)} versus the CN/\HCOP ratio. {\bf c)} The CS/HCN
ratio versus the CS/HNC ratio and {\bf d)} versus the CS/\HCOP
ratio. The symbols are described with Figure 1.}
\label{diagnostic2}
\end{center}
\end{figure*}

{\bf HCN, HNC and HCO$^+$ lines} - The observed ratios of HCN,
HNC, and \HCOP\ in Fig. \ref{FIR_hd_hd} show a range of values
between 0.2 and 2.0.  Modelling results indicate that the
\HCOP/HCN ratio in most PDR environments is expected to be smaller
than unity, while in XDRs the \HCOP/HCN line ratio becomes $\geq$
1 for the higher column densities.  The PDR sources would lie in
the lower parts ($\leq$ 1) of the upper two diagrams and in the
upper part ($\geq$ 1) of the bottom diagram of Figure
\ref{FIR_hd_hd}.

The observed HNC/HCN data shows significant clustering at 0.5
across the whole \irlum range, which could indicate that the
HNC/HCN ratios have been lowered by the evolving starburst
excitation. In the \HCOP/HCN diagram in Fig. \ref{FIR_hd_hd}b, the
OH MM sources are well confined to the 0.3 and 0.8 range with only
IC\,694 (Arp299A, IR\,11257+5850A) as the exception. Rather than
finding a steady decrease of the observed \HCOP/HCN with
increasing \irlum \citep{GraciaCarpioGPC2006}, our larger sample
presents a more complicated picture that hints at (systematic)
evolutionary influences on the nuclear medium.

The diagrams of Figure \ref{FIR_hd_hd} display systematic behavior
related to the \HCOP and HNC characteristics. The group of
megamaser sources (mostly) lies to the right of the vertical line
of \irlum $>$ 10$^{11}$ \lsol but they are in the PDR regions of
all three diagrams. They are mostly high $N_{\rm H}$ and pure PDR
sources. On the other hand, the group of sources left of that
vertical line lies in the PDR regime for the HNC/HCN ratio,
straddles the XDR-PDR dividing line for the \HCOP/HCN ratio, and
lies in the XDR region for the HNC/\HCOP ratio. This effect can be
attributed to \HCOP\ enhancement and HNC depletion during a
late-stage starburst phase, that pushes PDR sources to a first
order XDR regime in the two bottom diagrams. This late-PDR
behavior is exemplified by the late starburst M\,82.

The lowest values of HNC/\HCOP\ for high \irlum sources are for OH
MM IC\,694 and the two AGN sources NGC\,1068 (H$_2$O MM) and
NGC\,6240. NGC\,6240 is not an OH MM but is similar to Arp\,220 in
its properties. Both have dominant (circum-)nuclear starburst
activity and NGC\,6240 also has a shock-heated molecular emission
region located between its two AGN nuclei \citep{TacconiEA1999,
BaanHH2007, IonoEA2007}. The emission of the ring of starformation
in NGC\,1068 \citep{SchinnererEA2000} also falls within the SEST
observing beam; this would dominate the signature of the source
and results in different ratios than for IRAM 30m data
\citep{UseroEA2004}. As a result of these characteristics, the
three sources lie in the PDR region of Fig. \ref{FIR_hd_hd}a and
in the XDR region of Fig. \ref{FIR_hd_hd}c. Alternatively, the
emission from these sources could be dominated by low $N_{\rm H}$
emission regions, where the theoretical line ratios are reversed
compared to the high $N_{\rm H}$ cases (Sec. \ref{sec:pumping}).

The evolved starburst behavior displayed in Fig. \ref{FIR_hd_hd}
shows that the first-order PDR-XDR division of the three diagrams
may not be optimal. The most reliable tracer of PDR versus XDR
behavior appears to be the HNC/HCN ratio, because it is least
affected by late-stage starburst conditions. Using this ratio
would also imply that most sources of our sample are PDR dominated
starburst sources.

{\bf CN and CS lines} - The $N$(CN)/$N$(HCN) ratios would lie
below 2.0 for PDRs and (far) above for XDRs
\citep{MeijerinkSI2006}. In Figure \ref{FIR_cncs_hcn}ab any XDR
sources would lie (far) above the line in the two diagrams. The OH
MM have a typical CN(1$-$0)/HCN ratio in the 0.4--0.8 range,
except for Arp\,220 with a value of 1.7. The PDR-XDR division
appears well situated in the diagram.

The $N$(CS)/$N$(HCN) ratio should be $\geq$ 1 at high column
densities for both XDRs and PDRs (Fig. \ref{FIR_cncs_hcn}c). The
observed CS(3$-$2)/HCN(1$-$0) ratio appears to increase with
increasing \irlums, which suggests a correlation with column
density.

\begin{figure*}[htbp]
\begin{center}
\includegraphics[width=12cm,clip]{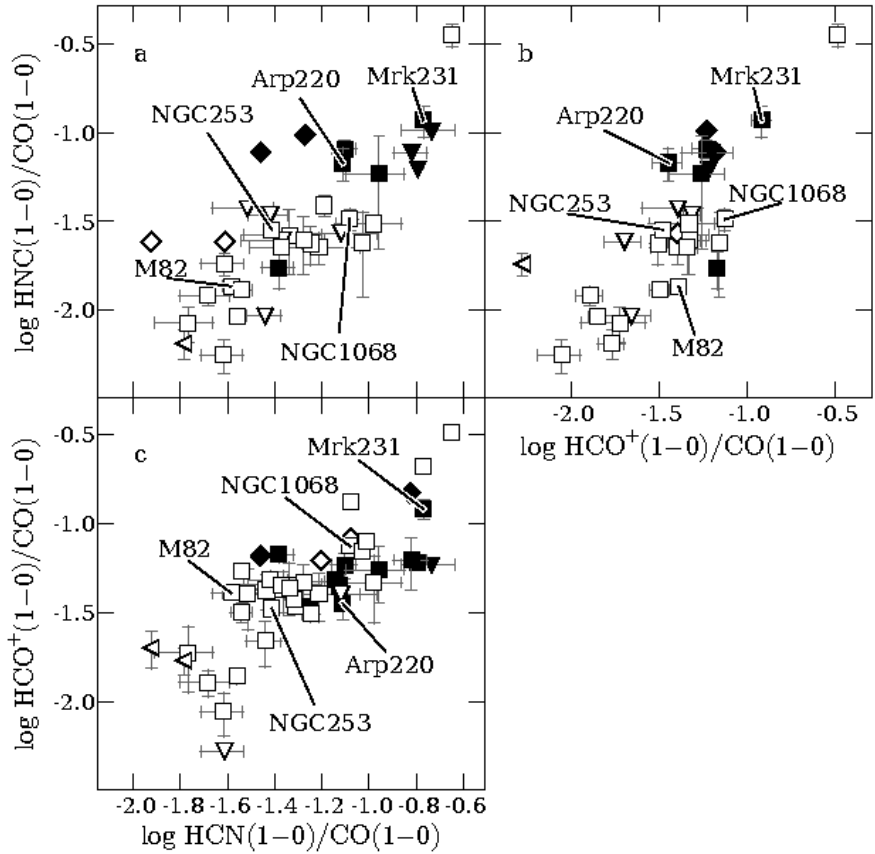}
\caption{Integrated line ratios of HCN\,(1$-$0), HNC\,(1$-$0), and
\HCOP\,(1$-$0) versus CO\,(1$-$0). The filled squares are values
for OH and \FORM\ megamaser galaxies, while the open squares are
values for mostly nearby starburst galaxies. {\bf a)} Integrated
HNC\,(1$-$0)/CO\,(1$-$0) versus HCN\,(1$-$0)/CO\,(1$-$0)\ ratios,
{\bf b)} integrated HNC\,(1$-$0)/CO\,(1$-$0) versus
\HCOP\,(1$-$0)/CO\,(1$-$0) ratios, and {\bf c)} integrated
\HCOP\,(1$-$0)/CO\,(1$-$0) versus HCN\,(1$-$0)/CO\,(1$-$0) ratios.
The symbols are described with Figure 1.} \label{diagnostic3}
\end{center}
\end{figure*}

\subsection{High-density tracer ratios versus each other}
\label{sec:diagnostics}

{\bf Tracers HCN, HCO$^+$, and HNC} - The line ratios of
high-density tracer molecules HCN, \HCOP, and HNC may be compared
to each other in order to discern collective behavior (Fig.
\ref{diagnostic1}). There is a significant spread in data points
and a separation of a group of mostly OH MM from an extended group
of non$-$OH MM. The discrepant OH MM point is again IC\,694.

The diagrams of Fig. \ref{diagnostic1} may be interpreted in the
framework of dominant PDR and XDR characteristics of the emission
regions using only the (more dependable) HNC/HCN ratio (see
Section \ref{sec:pumping}). The data points have extended
(slanted) distributions in all three frames with most sources
lying on the PDR side of the three dividing lines.

The dependence on density is strongest for \HCOP
\citep{MeijerinkSI2007}. In the top-left frame, the diagonal
structure represents a change from higher (bottom-right) to lower
(top-left) densities with Arp\,220 and other OH MM lying at the
high density extreme. HNC depletion and \HCOP enhancement due to
evolved starburst environmental effects (see Sec.
\ref{sec:pumping}) may account for the bunching of data points,
and the fact that some sources have moved close to the XDR-PDR
boundary.

{\bf Tracers CN and CS} - The CN/HNC and CN/HCO$^+$ versus CN/HCN
distributions show a significant spread with the OH MM
concentrated at lower CN/HNC ratios (Fig. \ref{diagnostic2}a). The
first order diagnostics discussed in Sec.\,8  indicates that all
sources have a CN/HCN ratio close to unity, which is consistent
with PDR environments. There is a hint of a gap in the source
distribution at CN/HCN $\approx$ 1. The slant in the data
distribution in the bottom frame of Fig. \ref{diagnostic2}a (as
compared with the top frame) results from the \HCOP\ enhancement
for part of the sample.

The CS diagrams (Fig. \ref{diagnostic2}b) display a clear extended
structure with the OH MM showing high ratios against HNC, HCO$^+$
and HCN. Since CS is to first order a column density tracer (see
Sec. \ref{sec:HCNexcitation} and \ref{sec:pumping}), this linear
variation in the relative strength of CS is a consequence of
seeing more of the highest density components with a higher column
density. ULIRGs/OH MM with high column densities have high ratios
and moderate (low) power FIR sources like NGC\,1808, M\,83, and
Maffei\,2 have smaller values.

\begin{figure}[htbp]
%\begin{center}
\includegraphics[width=7.5cm,clip]{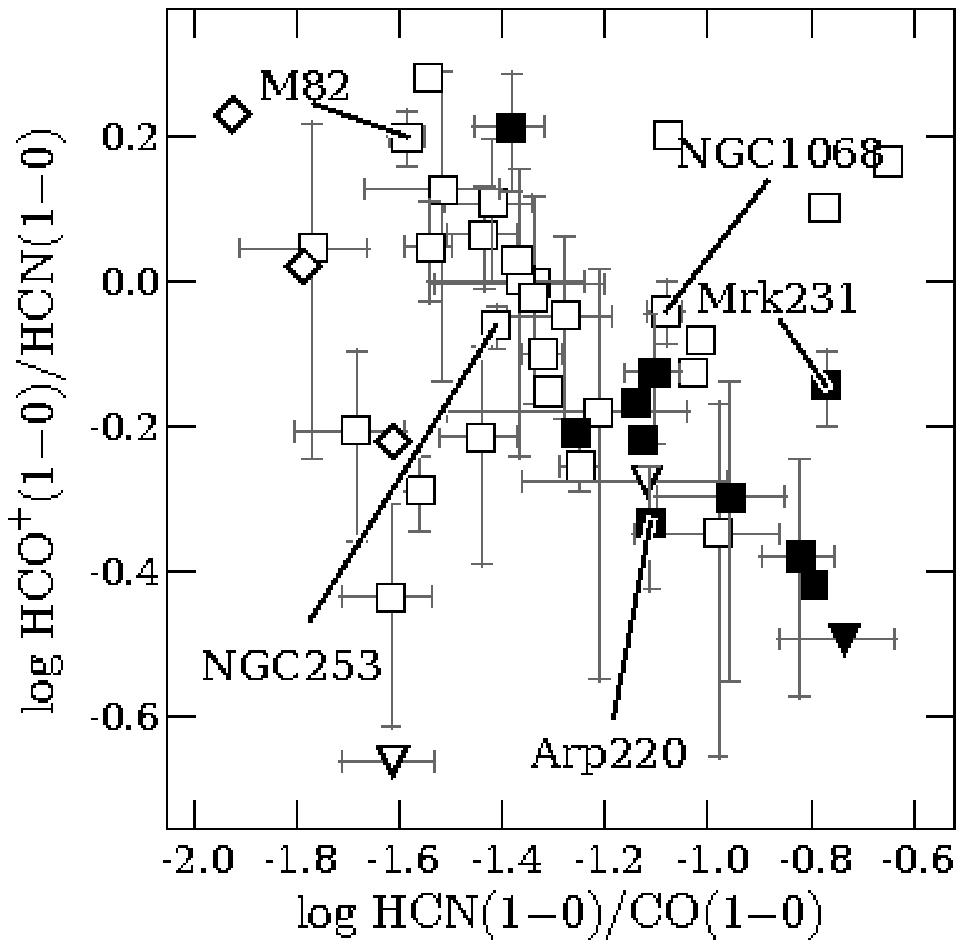}
\caption{The \HCOP\,(1$-$0)/HCN\,(1$-$0) ratio versus the
HCN\,(1$-$0)/CO\,(1$-$0) ratio. The symbols are described with
Figure 1.} \label{diagnostic4}
%\end{center}
\end{figure}

{\bf Comparisons with CO} - A diagnostic diagram may be
constructed using the ratios of tracers with CO\,(1$-$0) (Fig.
\ref{diagnostic3}). These diagrams also display a significant
(diagonal) spread in data points and a separation between OH MM
and the rest of the data points. The \HCOP enhancement and the
less apparent HNC depletion may account for the curved
distributions of the bottom-left and top-right frames. As
presented in previous sections, the ratios of high-density tracers
with CO would be evolutionary indicators that reduce in value
during the course of the outburst. Depending on their
\irlums(max), the sources during early evolutionary stages have a
starting point further towards the top-right side of each frame,
while more evolved sources or those with smaller \irlums(max) will
appear more towards the bottom-left side. The relative locations
of (late-stage) M\,82 and (early-stage low \irlums(max)) NGC\,253,
and ULIRGs/OH MM Arp\,220 and Mrk\,231 are consistent with this
picture. While there is no one-to-one interpretation of relative
positions of individual data points, the collective paths would
support FIR-related evolution from higher \irlum and
PDR$-$dominated sources in early stages of evolution to XDR$-$like
sources at lower \irlum (see Figs. \ref{diagnostic1},
\ref{diagnostic2}, \ref{diagnostic3}, and \ref{diagnostic4}).
Detailed modelling needs to confirm the evolutionary paths
followed by the nuclear activity having different initial
conditions.

\subsection{Diagnostics of the nuclear energy source}

The variation of the line intensity ratios has also been connected
to the AGN or starburst nucleus (SBN) nature of the nuclear power
sources \citep{KohnoEA2001, GraciaCarpioGPC2006}. Both authors
suggest that the data points with \HCOP/HCN $\le$ 0.55 would
indicate AGN activity and that higher values indicate a composite
AGN nature.

The (log-log) diagram of \HCOP/HCN versus the evolutionary
parameter of the HCN/CO ratio (Fig. \ref{diagnostic4}) shows a
continuous distribution for our larger sample (see Secs.
\ref{sec:HD/COevol} and \ref{sec:evolution}) that confirms the
distribution of data points presented earlier \citep{KohnoEA2001,
GraciaCarpioGPC2006}. The central ridge that accounts for most of
the data points shows an inverse-linear relation between the two
ratios, which suggests a general depletion of HCN relative to CO
and \HCOP\ during the course of the outburst. In our evolutionary
scenario, a source moves from (bottom) right to (upper) left in
the diagram during the decay segment of the nuclear outburst.
M\,82 represents a later stage of starburst evolution and is
located at the upper left, since its late-PDR nature also enhances
HCO$^+$. OH MMs and ULIRGs may start at the right part of Fig.
\ref{diagnostic4}, while less dominant starbursts have a starting
point more at the center or even on the left hand side of the
diagram. NGC\,253 is representing such a case.

First order diagnostics (Sec. \ref{sec:pumping}) relates the
\HCOP/HCN ratio to PDR and XDR signatures rather than to the
nature of central power source. We hereby assume that high $N_{\rm
H}$ PDRs dominate the emission signatures and produce the low
values for the \HCOP/HCN ratio, rather than low $N_{\rm H}$ XDRs.
Of the four OH MM sources with the lowest ($\le$ 0.55 or log $\le$
-0.26) ratios (Arp\,220, IRAS\,11506-3851, IRAS\,15107+0724, and
Mrk\,273), the radio classification only indicates a possible AGN
in IRAS\,15107+0724 \citep{BaanKlockner2006}. NIR classification
suggests an AGN power contribution of 50\% in Mrk\,273
\citep{GenzelEA1998}.  Both these classifications suggest a
significant AGN presence in Mrk\,231, which has a ratio here of
0.72 \cite[higher than in e.g.][]{GraciaCarpioGPC2006}.  Observed
\HCOP/HCN values are also higher than 0.55 for the starburst plus
AGN source NGC\,1068 \citep[0.9;][]{UseroEA2004} and the nearby
starburst-dominated sources NGC\,253 (0.9) and M\,82
\citep[1.6;][]{FuenteEA2005}.

Therefore, the predicted properties of the nuclear ISM during the
evolution of the starburst may well account for the data points in
Fig. \ref{diagnostic4} and particularly the (inverse-linear)
central ridge. If the observed trend were affected by the presence
of an AGN, its influence on the excitation environment of the
nucleus would have to be further confirmed by more detailed
modelling. The (circum-) nuclear starburst would easily outshine
the compact AGN region and dominate the emission signature in low
spatial resolution data.

\section{Summary}

Molecular line emissions in FIR-luminous galaxies are tools for
multi-dimensional diagnostics of the environmental parameters in
the nuclear ISM and the heating processes resulting from the
nuclear activity.  A total of 118 line detections made with the
30-m Pico Veleta and the 15-m SEST telescopes for a total of 37
sources have been complemented with partial records published for
another 80 sources.

A proper evaluation of integrated emission lines and their
diagnostic line ratios can be achieved after synthesis and
modelling of a representative description of the integrated
nuclear molecular medium under a wide variety of AGN- and
starburst-related circumstances. The molecular information in this
paper has been diagnosed to first order using modelling results
for nuclear emission regions presented in the literature.

The collective behavior of line luminosities and line ratios of
the low-density and high-density tracers presents a consistent
picture of the molecular medium in the nuclear regions of ULIRGs.
The high-density tracers represent the molecular medium in the
regions where star formation is taking place, and the low-density
tracer represents the relatively unperturbed larger-scale
molecular environment. The tracer luminosities increase roughly
linearly with FIR luminosity but they show significant scatter in
data points due to physical processes in the nuclear region, the
chemical history, and the relative age of the nuclear activity.
The luminosity of high-density tracers varies linearly with \irlum
except for the CS(3$-$2) luminosity, where a steeper relation is
suggestive of a different excitation dependence. The CO(1$-$0) and
CO(2$-$1) lines for these high FIR luminosity sources have a slope
less than unity and are (almost) consistent with empirical
evidence on the relation of SFR versus gas content
\citep{Kennicutt1998}.

First order diagnostics of the nuclear ISM can be based on the
collective behavior of high-density tracers HCN, HNC and \HCOP\
showing significant differentiation and systematic changes. The
line strengths of these three molecules relative to the
CO\,(1$-$0) line show a significant range of 0.03 to 0.23 with a
distinct dependence on \irlums. This dependence has been
interpreted in terms of an evolutionary model where the depletion
of molecular gas depends on the consumption and destruction of the
high-density gas by the ongoing star formation process. Only
partial diagnostics could be done using the available CN and CS
data, which complements the diagnostics of the three other
high-density tracers.

The emission line ratios of the three high-density tracers of the
galaxies show a structured distribution that fills selected parts
of the parameter space. Interpreting these distributions has been
done using modelling column densities that are characteristic of
PDR- and XDR-dominated nuclear environments. OH MM and other
powerful ULIRGs are mostly characterized by PDR-dominance. The
other end of the data distribution is characterized by mixed
PDR-XDR and XDR-dominated environments. (U)LIRGs represent the
phases of nuclear evolution that generate the highest FIR
luminosities and most rapidly deplete the dense molecular
component. Using the HCN/CO ratio as an evolutionary indicator,
the distributions of data points may reflect evolution from
PDR-like to more XDR-like nuclear ISM properties during the course
of the outburst.

The \HCOP/HCN and HNC/HCN ratios serve as indicators of
environments with shocks and non-standard heating and with HNC
depletion and \HCOP\ enhancement in a fraction of the sources. The
\HCOP/HCN ratio also serves as a density indicator and would be
higher under PDR circumstances and lower in XDR environments. The
observed trends in the \HCOP/HCN ratio could result from the
(indirect) influence of an AGN in the nucleus. More likely the
trends in the relative abundance of HCN as compared with other
constituents result from evolution of the nuclear environment
during the SBN activity.

The detailed interpretations of multi-line multi-molecule emission
line behavior and their relation with the local environment
require detailed modelling of the physical parameters of the
environment together with the excitation, the chemistry, and the
radiative transfer for the molecular constituents. Hereby the
integration of higher level transitions of the molecules is needed
to determine specific excitation temperatures and densities. A
comparison with the properties of Galactic emission regions will
emphasize the effect of scale size on the line ratios for tracer
molecules. Further studies are underway to connect the FIR
signature and global heating scenario of ULIRGs
\citep{LoenenBS2006} with modelling of the molecular emissions
under varying conditions in a nuclear starburst. The emission
scenarios for XDRs and PDRs establish the connection between star
formation and other sources of excitation and the integrated
emission line parameters observed in extragalactic sources.

\bigskip

\begin{acknowledgements} The authors thank the IRAM staff at
Pico Veleta and the staff of the Onsala Space Observatory at the
SEST telescope for their valuable help during the observations.
WAB thanks Wolfram Freudling for support during the startup of the
SEST project and Lydia Haitas Baan for support with the
observations during the second SEST run. WAB also thanks Arecibo
REU summer student Melissa Nysewander (Univ. of North Carolina)
for support during data reduction. The authors thank Marco Spaans
for consultations on modelling aspects. This research has made use
of the NASA/IPAC Extragalactic Database (NED), which is operated
by the Jet Propulsion Laboratory, Caltech, under contract with
NASA. This research has also made use of NASA's Astrophysics Data
System Abstract Service. The authors thank the referee Melanie
Krips for constructive comments and critical reading of the paper.
\end{acknowledgements}

\bibliographystyle{aa}
\bibliography{APMolecV3}

\begin{thebibliography}{70}
\expandafter\ifx\csname natexlab\endcsname\relax\def\natexlab#1{#1}\fi

\bibitem[{{Aalto} {et~al.}(2002){Aalto}, {Polatidis}, {H{\"u}ttemeister}, \&
  {Curran}}]{AaltoPHC2002}
{Aalto}, S., {Polatidis}, A.~G., {H{\"u}ttemeister}, S., \& {Curran}, S.~J.
  2002, \aap, 381, 783

\bibitem[{{Aalto} {et~al.}(2007){Aalto}, {Spaans}, {Wiedner}, \&
  {H{\"u}ttemeister}}]{AaltoEA2007}
{Aalto}, S., {Spaans}, M., {Wiedner}, M.~C., \& {H{\"u}ttemeister}, S. 2007,
  \aap, 464, 193

\bibitem[{{Araya} {et~al.}(2004){Araya}, {Baan}, \& {Hofner}}]{ArayaBH2004}
{Araya}, E., {Baan}, W.~A., \& {Hofner}, P. 2004, \apjs, 154, 541

\bibitem[{{Baan}(1989)}]{Baan1989}
{Baan}, W.~A. 1989, \apj, 338, 804

\bibitem[{{Baan} {et~al.}(2007){Baan}, {Hagiwara}, \& {Hofner}}]{BaanHH2007}
{Baan}, W.~A., {Hagiwara}, Y., \& {Hofner}, P. 2007, \apj, 661, 173

\bibitem[{{Baan} {et~al.}(1993){Baan}, {Haschick}, \& {Uglesich}}]{BaanHU1993}
{Baan}, W.~A., {Haschick}, A.~D., \& {Uglesich}, R. 1993, \apj, 415, 140

\bibitem[{{Baan} \& {Kl{\"o}ckner}(2006)}]{BaanKlockner2006}
{Baan}, W.~A. \& {Kl{\"o}ckner}, H.-R. 2006, \aap, 449, 559

\bibitem[{{Bally} {et~al.}(1988){Bally}, {Stark}, {Wilson}, \&
  {Henkel}}]{BallySWH1988}
{Bally}, J., {Stark}, A.~A., {Wilson}, R.~W., \& {Henkel}, C. 1988, \apj, 324,
  223

\bibitem[{{Bevington} \& {Robinson}(1992)}]{Bevington1992}
{Bevington}, P.~R. \& {Robinson}, D.~K. 1992, {Data reduction and error
  analysis for the physical sciences} (New York: McGraw-Hill, |c1992, 2nd ed.)

\bibitem[{{Boger} \& {Sternberg}(2005)}]{BogerS2005}
{Boger}, G.~I. \& {Sternberg}, A. 2005, \apj, 632, 302

\bibitem[{{Braine} \& {Combes}(1992)}]{BraineCombes1992}
{Braine}, J. \& {Combes}, F. 1992, \aap, 264, 433

\bibitem[{{Churchwell} {et~al.}(1984){Churchwell}, {Nash}, \&
  {Walmsley}}]{1984ApJ...287..681C}
{Churchwell}, E., {Nash}, A.~G., \& {Walmsley}, C.~M. 1984, \apj, 287, 681

\bibitem[{{Dahmen} {et~al.}(1998){Dahmen}, {Huttemeister}, {Wilson}, \&
  {Mauersberger}}]{DahmenEA1998}
{Dahmen}, G., {Huttemeister}, S., {Wilson}, T.~L., \& {Mauersberger}, R. 1998,
  \aap, 331, 959

\bibitem[{{Dickinson} {et~al.}(1980){Dickinson}, {Dinger}, {Kuiper}, \&
  {Rodriguez Kuiper}}]{DickinsonEA1980}
{Dickinson}, D.~F., {Dinger}, A.~S.~C., {Kuiper}, T.~B.~H., \& {Rodriguez
  Kuiper}, E.~N. 1980, \apjl, 237, L43

\bibitem[{{Downes}(1989)}]{1989LNP...333..351D}
{Downes}, D. 1989, LNP Vol.~333: Evolution of Galaxies: Astronomical
  Observations, 333, 351

\bibitem[{{Downes} \& {Solomon}(1998)}]{DownesSolomon1998}
{Downes}, D. \& {Solomon}, P.~M. 1998, \apj, 507, 615

\bibitem[{{Elitzur}(1983)}]{Elitzur1983}
{Elitzur}, M. 1983, \apj, 267, 174

\bibitem[{{Fuente} {et~al.}(2005){Fuente}, {Garc{\'{\i}}a-Burillo}, {Gerin},
  {Teyssier}, {Usero}, {Rizzo}, \& {de Vicente}}]{FuenteEA2005}
{Fuente}, A., {Garc{\'{\i}}a-Burillo}, S., {Gerin}, M., {et~al.} 2005, \apjl,
  619, L155

\bibitem[{{Gao} \& {Solomon}(2004{\natexlab{a}})}]{GaoS2004a}
{Gao}, Y. \& {Solomon}, P.~M. 2004{\natexlab{a}}, \apjs, 152, 63

\bibitem[{{Gao} \& {Solomon}(2004{\natexlab{b}})}]{GaoS2004b}
{Gao}, Y. \& {Solomon}, P.~M. 2004{\natexlab{b}}, \apj, 606, 271

\bibitem[{{Garay} {et~al.}(1993){Garay}, {Mardones}, \&
  {Mirabel}}]{GarayMM1993}
{Garay}, G., {Mardones}, D., \& {Mirabel}, I.~F. 1993, \aap, 277, 405

\bibitem[{{Garc{\'{\i}}a-Burillo} {et~al.}(2002){Garc{\'{\i}}a-Burillo},
  {Mart{\'{\i}}n-Pintado}, {Fuente}, {Usero}, \& {Neri}}]{GarciaBEA2002}
{Garc{\'{\i}}a-Burillo}, S., {Mart{\'{\i}}n-Pintado}, J., {Fuente}, A.,
  {Usero}, A., \& {Neri}, R. 2002, \apjl, 575, L55

\bibitem[{{Genzel} {et~al.}(1998){Genzel}, {Lutz}, {Sturm}, {Egami}, {Kunze},
  {Moorwood}, {Rigopoulou}, {Spoon}, {Sternberg}, {Tacconi-Garman}, {Tacconi},
  \& {Thatte}}]{GenzelEA1998}
{Genzel}, R., {Lutz}, D., {Sturm}, E., {et~al.} 1998, \apj, 498, 579

\bibitem[{{Goldsmith} {et~al.}(1981){Goldsmith}, {Langer}, {Ellder},
  {Kollberg}, \& {Irvine}}]{GoldsmithEA1981}
{Goldsmith}, P.~F., {Langer}, W.~D., {Ellder}, J., {Kollberg}, E., \& {Irvine},
  W. 1981, \apj, 249, 524

\bibitem[{{Graci{\'a}-Carpio} {et~al.}(2006){Graci{\'a}-Carpio},
  {Garc{\'{\i}}a-Burillo}, {Planesas}, \& {Colina}}]{GraciaCarpioGPC2006}
{Graci{\'a}-Carpio}, J., {Garc{\'{\i}}a-Burillo}, S., {Planesas}, P., \&
  {Colina}, L. 2006, \apjl, 640, L135

\bibitem[{{Graedel} {et~al.}(1982){Graedel}, {Langer}, \&
  {Frerking}}]{GreadelLF1982}
{Graedel}, T.~E., {Langer}, W.~D., \& {Frerking}, M.~A. 1982, \apjs, 48, 321

\bibitem[{{Greaves} \& {Church}(1996)}]{GreavesC1996}
{Greaves}, J.~S. \& {Church}, S.~E. 1996, \mnras, 283, 1179

\bibitem[{{Greve} {et~al.}(2006){Greve}, {Papadopoulos}, {Gao}, \&
  {Radford}}]{GrevePGR2006}
{Greve}, T.~R., {Papadopoulos}, P.~P., {Gao}, Y., \& {Radford}, S.~J.~E. 2006,
  ArXiv Astrophysics e-prints

\bibitem[{{Henkel} {et~al.}(1991){Henkel}, {Baan}, \&
  {Mauersberger}}]{1991A&ARv...3...47H}
{Henkel}, C., {Baan}, W.~A., \& {Mauersberger}, R. 1991, \aapr, 3, 47

\bibitem[{{Henkel} \& {Wilson}(1990)}]{HenkelWilson1990}
{Henkel}, C. \& {Wilson}, T.~L. 1990, \aap, 229, 431

\bibitem[{{H\"uttemeister} {et~al.}(1995){H\"uttemeister}, {Henkel},
  {Mauersberger}, {Brouillet}, {Wiklind}, \& {Millar}}]{HuttemeisterEA1995}
{H\"uttemeister}, S., {Henkel}, C., {Mauersberger}, R., {et~al.} 1995, \aap,
  295, 571

\bibitem[{{Iono} {et~al.}(2007){Iono}, {Wilson}, {Takakuwa}, {Yun}, {Petitpas},
  {Peck}, {Ho}, {Matsushita}, {Pihlstrom}, \& {Wang}}]{IonoEA2007}
{Iono}, D., {Wilson}, C.~D., {Takakuwa}, S., {et~al.} 2007, \apj, 659, 283

\bibitem[{{Irvine} {et~al.}(1987){Irvine}, {Goldsmith}, \&
  {Hjalmarson}}]{1987ip...symp..561I}
{Irvine}, W.~M., {Goldsmith}, P.~F., \& {Hjalmarson}, A. 1987, in ASSL Vol.
  134: Interstellar Processes, ed. D.~J. {Hollenbach} \& H.~A. {Thronson}, Jr.,
  561--609

\bibitem[{{Jansen} {et~al.}(1995{\natexlab{a}}){Jansen}, {Spaans},
  {Hogerheijde}, \& {van Dishoeck}}]{JansenSHD1995}
{Jansen}, D.~J., {Spaans}, M., {Hogerheijde}, M.~R., \& {van Dishoeck}, E.~F.
  1995{\natexlab{a}}, \aap, 303, 541

\bibitem[{{Jansen} {et~al.}(1995{\natexlab{b}}){Jansen}, {van Dishoeck},
  {Black}, {Spaans}, \& {Sosin}}]{JansenEA1995}
{Jansen}, D.~J., {van Dishoeck}, E.~F., {Black}, J.~H., {Spaans}, M., \&
  {Sosin}, C. 1995{\natexlab{b}}, \aap, 302, 223

\bibitem[{{Kennicutt}(1998)}]{Kennicutt1998}
{Kennicutt}, Jr., R.~C. 1998, \apj, 498, 541

\bibitem[{{Kohno} {et~al.}(2001){Kohno}, {Matsushita}, {Vila-Vilar{\'o}},
  {Okumura}, {Shibatsuka}, {Okiura}, {Ishizuki}, \& {Kawabe}}]{KohnoEA2001}
{Kohno}, K., {Matsushita}, S., {Vila-Vilar{\'o}}, B., {et~al.} 2001, in ASP
  Conf. Ser. 249: The Central Kiloparsec of Starbursts and AGN: The La Palma
  Connection, ed. J.~H. {Knapen}, J.~E. {Beckman}, I.~{Shlosman}, \& T.~J.
  {Mahoney}, 672

\bibitem[{{Krolik} \& {Kallman}(1983)}]{1983ApJ...267..610K}
{Krolik}, J.~H. \& {Kallman}, T.~R. 1983, \apj, 267, 610

\bibitem[{{Krumholz} \& {Thompson}(2007)}]{KrumholzThompson2007}
{Krumholz}, M.~R. \& {Thompson}, T.~A. 2007, ArXiv e-prints, 704

\bibitem[{{Lepp} \& {Dalgarno}(1996)}]{1996A&A...306L..21L}
{Lepp}, S. \& {Dalgarno}, A. 1996, \aap, 306, L21

\bibitem[{{Loenen} {et~al.}(2006){Loenen}, {Baan}, \& {Spaans}}]{LoenenBS2006}
{Loenen}, A.~F., {Baan}, W.~A., \& {Spaans}, M. 2006, \aap, 458, 89

\bibitem[{{Maloney} \& {Black}(1988)}]{MaloneyB1988}
{Maloney}, P. \& {Black}, J.~H. 1988, \apj, 325, 389

\bibitem[{{Mauersberger} \& {Henkel}(1989)}]{MauersbergerH1989}
{Mauersberger}, R. \& {Henkel}, C. 1989, \aap, 223, 79

\bibitem[{{Mauersberger} {et~al.}(1996{\natexlab{a}}){Mauersberger}, {Henkel},
  {Whiteoak}, {Chin}, \& {Tieftrunk}}]{MauersbergerEA1996b}
{Mauersberger}, R., {Henkel}, C., {Whiteoak}, J.~B., {Chin}, Y.-N., \&
  {Tieftrunk}, A.~R. 1996{\natexlab{a}}, \aap, 309, 705

\bibitem[{{Mauersberger} {et~al.}(1996{\natexlab{b}}){Mauersberger}, {Henkel},
  {Wielebinski}, {Wiklind}, \& {Reuter}}]{MauersbergerEA1996a}
{Mauersberger}, R., {Henkel}, C., {Wielebinski}, R., {Wiklind}, T., \&
  {Reuter}, H.-P. 1996{\natexlab{b}}, \aap, 305, 421

\bibitem[{{Mauersberger} {et~al.}(1989){Mauersberger}, {Henkel}, {Wilson}, \&
  {Harju}}]{MauersbergerHWH1989}
{Mauersberger}, R., {Henkel}, C., {Wilson}, T.~L., \& {Harju}, J. 1989, \aap,
  226, L5

\bibitem[{{McQuinn} {et~al.}(2002){McQuinn}, {Simon}, {Law}, {Jackson},
  {Bania}, {Clemens}, \& {Heyer}}]{McQuinnEA2002}
{McQuinn}, K.~B.~W., {Simon}, R., {Law}, C.~J., {et~al.} 2002, \apj, 576, 274

\bibitem[{{Meijerink} \& {Spaans}(2005)}]{MeijerinkS2005}
{Meijerink}, R. \& {Spaans}, M. 2005, \aap, 436, 397

\bibitem[{{Meijerink} {et~al.}(2006){Meijerink}, {Spaans}, \&
  {Israel}}]{MeijerinkSI2006}
{Meijerink}, R., {Spaans}, M., \& {Israel}, F.~P. 2006, \apjl, 650, L103

\bibitem[{{Meijerink} {et~al.}(2007){Meijerink}, {Spaans}, \&
  {Israel}}]{MeijerinkSI2007}
{Meijerink}, R., {Spaans}, M., \& {Israel}, F.~P. 2007, \aap, 461, 793

\bibitem[{{Mirabel} {et~al.}(1990){Mirabel}, {Booth}, {Johansson}, {Garay}, \&
  {Sanders}}]{MirabelEA1990}
{Mirabel}, I.~F., {Booth}, R.~S., {Johansson}, L.~E.~B., {Garay}, G., \&
  {Sanders}, D.~B. 1990, \aap, 236, 327

\bibitem[{{Nguyen-Q-Rieu} {et~al.}(1992){Nguyen-Q-Rieu}, {Jackson}, {Henkel},
  {Truong}, \& {Mauersberger}}]{NguyenEA1992}
{Nguyen-Q-Rieu}, {Jackson}, J.~M., {Henkel}, C., {Truong}, B., \&
  {Mauersberger}, R. 1992, \apj, 399, 521

\bibitem[{{Papadopoulos}(2007)}]{Papadop2007}
{Papadopoulos}, P.~P. 2007, \apj, 656, 792

\bibitem[{{Radford} {et~al.}(1991){Radford}, {Downes}, \&
  {Solomon}}]{RadfordDS1991}
{Radford}, S.~J.~E., {Downes}, D., \& {Solomon}, P.~M. 1991, \apjl, 368, L15

\bibitem[{{Rodriguez-Franco} {et~al.}(1998){Rodriguez-Franco},
  {Martin-Pintado}, \& {Fuente}}]{RodriguezEA1998}
{Rodriguez-Franco}, A., {Martin-Pintado}, J., \& {Fuente}, A. 1998, \aap, 329,
  1097

\bibitem[{{Sanders} \& {Mirabel}(1996)}]{SandersM1996}
{Sanders}, D.~B. \& {Mirabel}, I.~F. 1996, \araa, 34, 749

\bibitem[{{Schilke} {et~al.}(1992){Schilke}, {Walmsley}, {Pineau Des Forets},
  {Roueff}, {Flower}, \& {Guilloteau}}]{SchilkeEA1992}
{Schilke}, P., {Walmsley}, C.~M., {Pineau Des Forets}, G., {et~al.} 1992, \aap,
  256, 595

\bibitem[{{Schinnerer} {et~al.}(2000){Schinnerer}, {Eckart}, {Tacconi},
  {Genzel}, \& {Downes}}]{SchinnererEA2000}
{Schinnerer}, E., {Eckart}, A., {Tacconi}, L.~J., {Genzel}, R., \& {Downes}, D.
  2000, \apj, 533, 850

\bibitem[{{Snell} {et~al.}(1984){Snell}, {Goldsmith}, {Erickson}, {Mundy}, \&
  {Evans}}]{SnellEA1984}
{Snell}, R.~L., {Goldsmith}, P.~F., {Erickson}, N.~R., {Mundy}, L.~G., \&
  {Evans}, II, N.~J. 1984, \apj, 276, 625

\bibitem[{{Sodroski} {et~al.}(1995){Sodroski}, {Odegard}, {Dwek}, {Hauser},
  {Franz}, {Freedman}, {Kelsall}, {Wall}, {Berriman}, {Odenwald}, {Bennett},
  {Reach}, \& {Weiland}}]{SodroskiEA1995}
{Sodroski}, T.~J., {Odegard}, N., {Dwek}, E., {et~al.} 1995, \apj, 452, 262

\bibitem[{{Solomon} {et~al.}(1997){Solomon}, {Downes}, {Radford}, \&
  {Barrett}}]{SolomonEA1997}
{Solomon}, P.~M., {Downes}, D., {Radford}, S.~J.~E., \& {Barrett}, J.~W. 1997,
  \apj, 478, 144

\bibitem[{{Spaans} \& {Meijerink}(2007)}]{SpaansM2007}
{Spaans}, M. \& {Meijerink}, R. 2007, \apjl, 664, L23

\bibitem[{{Strong} {et~al.}(1988){Strong}, {Bloemen}, {Dame}, {Grenier},
  {Hermsen}, {Lebrun}, {Nyman}, {Pollock}, \& {Thaddeus}}]{StrongEA1988}
{Strong}, A.~W., {Bloemen}, J.~B.~G.~M., {Dame}, T.~M., {et~al.} 1988, \aap,
  207, 1

\bibitem[{{Tacconi} {et~al.}(1999){Tacconi}, {Genzel}, {Tecza}, {Gallimore},
  {Downes}, \& {Scoville}}]{TacconiEA1999}
{Tacconi}, L.~J., {Genzel}, R., {Tecza}, M., {et~al.} 1999, \apj, 524, 732

\bibitem[{{Ungerechts} {et~al.}(1997){Ungerechts}, {Bergin}, {Goldsmith},
  {Irvine}, {Schloerb}, \& {Snell}}]{1997ApJ...482..245U}
{Ungerechts}, H., {Bergin}, E.~A., {Goldsmith}, P.~F., {et~al.} 1997, \apj,
  482, 245

\bibitem[{{Usero} {et~al.}(2004){Usero}, {Garc{\'{\i}}a-Burillo}, {Fuente},
  {Mart{\'{\i}}n-Pintado}, \& {Rodr{\'{\i}}guez-Fern{\'a}ndez}}]{UseroEA2004}
{Usero}, A., {Garc{\'{\i}}a-Burillo}, S., {Fuente}, A.,
  {Mart{\'{\i}}n-Pintado}, J., \& {Rodr{\'{\i}}guez-Fern{\'a}ndez}, N.~J. 2004,
  \aap, 419, 897

\bibitem[{{Wang} {et~al.}(2004){Wang}, {Henkel}, {Chin}, {Whiteoak}, {Hunt
  Cunningham}, {Mauersberger}, \& {Muders}}]{WangEA2004}
{Wang}, M., {Henkel}, C., {Chin}, Y.-N., {et~al.} 2004, \aap, 422, 883

\bibitem[{{Wootten}(1981)}]{Wootten1981}
{Wootten}, A. 1981, \apj, 245, 105

\bibitem[{{Wu} {et~al.}(2005){Wu}, {Evans}, {Gao}, {Solomon}, {Shirley}, \&
  {Vanden Bout}}]{WuEvansEA2005}
{Wu}, J., {Evans}, II, N.~J., {Gao}, Y., {et~al.} 2005, \apjl, 635, L173

\bibitem[{{Young} \& {Scoville}(1991)}]{YoungScoville1991}
{Young}, J.~S. \& {Scoville}, N.~Z. 1991, \araa, 29, 581

\end{thebibliography}
%
% --------------------------------------------------------------
% ONLINE Material
%---------------------------------------------------------------
%

\vspace*{2cm} The Online sections of this paper are not presented
here. The complete paper can be found at
www.astron.nl/$\sim$willem/paper\_links.html

\end{document}